\documentclass[aps,prd,showpacs,eqsecnum,notitlepage,nofootinbib]{revtex4-1}
\usepackage[centertags]{amsmath}
\usepackage{amssymb}
\usepackage{amsfonts,mathrsfs}
\usepackage{hyperref}	
\usepackage{graphicx}
\usepackage{dcolumn}% Align table columns on decimal point
\usepackage{bm}% bold math
\usepackage{tabularx}
\usepackage{array}
\usepackage{float}

\newcommand{\bo}{\raise-1mm\hbox{\Large$\Box$}}

\newcommand{\f}[2]{\frac{#1}{#2}}

\newcommand{\la}{\langle}
\newcommand{\ra}{\rangle}
\newcommand{\w}{\omega}
\newcommand{\kp}{\kappa}
\newcommand{\be}{\begin{equation}}
\newcommand{\ee}{\end{equation}}
\newcommand{\bea}{\begin{eqnarray}}
\newcommand{\eea}{\end{eqnarray}}
\newcommand{\bes}{\begin{subequations}}
\newcommand{\ese}{\end{subequations}}
\newcommand{\eqn}[1]{(\ref{#1})}

\begin{document}

\title{Time Dependence of Particle Creation from Accelerating Mirrors}
\author{Michael R.R. Good}
\email{mgood@ntu.edu.sg}
\affiliation{Institute of Advanced Studies, Nanyang Technological University,
Singapore.}

\author{Paul R. Anderson}
\email{anderson@wfu.edu}
\affiliation{Department of Physics, Wake Forest University,
Winston-Salem, North Carolina, 27109,
USA.}

\author{Charles R. Evans}
\email{evans@physics.unc.edu}
\affiliation{Department of Physics and Astronomy, University of North
Carolina, Chapel Hill, North Carolina, 27599, USA.}

\date{\today}

\begin{abstract}

Particle production due to a quantized, massless, minimally coupled scalar
field in two-dimensional flat spacetime with an accelerating mirror is
investigated, with a focus on the time dependence of the process.  We analyze
first the classes of trajectories previously investigated by Carlitz and
Willey and by Walker and Davies.  We then analyze four new classes of
trajectories, all of which can be expressed analytically and for which
several ancillary properties can be derived analytically.  The time
dependence is investigated through the use of wave packets for the modes of
the quantized field that are in the \textit{out} vacuum state.
It is shown for most of the trajectories studied that good
time resolution of the particle production process can be obtained.

\end{abstract}


\maketitle

\section{\label{sec:level1}INTRODUCTION}

As the simplest theoretical manifestation of the dynamical Casimir effect
(DCE), the moving mirror model of DeWitt~\cite{DeWitt:1975ys}, and Davies and
Fulling~\cite{Davies:1976hi,Davies:1977yv} describes the
disturbance of a field by an accelerated boundary, which results in both
particle production and a flux of energy.  As the mirror model
matured~\cite{Candelas:1977zza,Ford:1982ct,Unruh:1982ic,Carlitz:1986nh,
Golestanian:1998bx}, it became apparent that accelerating boundaries could be
used to understand entropy production \cite{Mukohyama:2000tv,Page:2000ni},
the relationship between particles and energy \cite{Walker:1984vj}, and
thermodynamical paradoxes \cite{Helfer:2000fg,Walker:1984ya,Davies:1982cn}.
The DCE has the potential to be measured
\cite{Braggio:2005,PhysRevLett.96.200402}.  Indeed, one experiment claims
to have effectively measured the DCE using a superconducting quantum
interference device (SQUID) that acts as a moving mirror \cite{BOOYAA}.  In
another recent set of experiments~\cite{Jaskula:2012} an analogue of the DCE
has been observed in the case of a Bose-Einstein condensate.

One aspect of the moving mirror model that has been largely unexplored is
the study of the time dependence of the particle production process.  The
time dependence of the stress-energy tensor for the quantized field has been
worked out.  However, the stress-energy contains vacuum polarization effects
along with particle production effects, and in most cases there is no clear
way of separating the two.  In a flat-space background the Bogolubov
transformation between the \textit{in} and \textit{out} vacuum states can
be used to accurately describe the particle production process.  However,
the resulting particle frequency spectrum retains no information regarding
the time dependence of creation.

One way to explore the time dependence of particle production is through the
use of wave packets.  Hawking~\cite{Hawking:1974sw} made use of such packets
to describe the late-time behavior of black hole radiation when a black hole
forms from collapse.  In the context of a moving mirror, wave packets have
been used by Dorca and Verdaguer~\cite{Dorca:1993si} for a specific class
of trajectories that generate a thermal spectrum at late times.

In this paper we explore the use of wave packets as a means of obtaining
information about the time dependence of particle production due to
accelerating mirrors in (1+1) dimensions.  We do so using two previously
investigated trajectories and four new ones.  In each case we compare the
particle creation results with the stress-energy of the quantum field.  We
work with a massless minimally coupled free scalar field.  The packets are
obtained by integrating the
modes of the quantum field over specific frequency ranges using a
parameterized weighting function that creates a packet whose amplitude is
largest near a particular time that is related to the value of one of the
parameters.  By computing the Bogolubov transformation
using packets for the modes in the \textit{out} vacuum state it is possible
to obtain an expression for the number of particles produced in various
frequency and time intervals as a function of time~\cite{Fabbri:2005mw}.
There is a fundamental uncertainty principle involved in working with the
wave packets such that small frequency bins lead to good resolution in
frequency and poor resolution in time, and vice versa.  We explore the
effects of this uncertainty relation for the trajectories chosen.

To facilitate the investigation we have restricted our attention to mirror
trajectories for which the Bogolubov components can be computed analytically.
Only a small number of classes of trajectories have previously been considered
for which it is possible to analytically compute the Bogolubov coefficients.
These include the trajectories studied by Carlitz and
Willey~\cite{Carlitz:1986nh}, which are designed so that there is constant
flux of energy, and the class of trajectories studied
by Walker and Davies~\cite{Walker:1982}, which involve a mirror that starts
at rest in the infinite past, accelerates, and ends at rest in the infinite
future.  In this paper we introduce four new trajectories for which the
Bogolubov coefficients can be computed analytically.  Three of these involve
mirrors that start at rest in the infinite past.  In one case the mirror
trajectory
is also asymptotically static in the future, in another the mirror's speed is
asymptotically constant, and in the third the mirror's speed approaches the
speed of light.  In this third case the trajectory nevertheless ends up at
future timelike infinity instead of being asymptotic to a null trajectory.
A fourth class of trajectories begins in the same way as those studied by
Carlitz and Willey but then, instead of becoming asymptotic to a null ray at
late times, approaches a constant velocity and thus becomes inertial.

In the only previous use of wave packets for the moving mirror model that we
are aware of, Dorca and Verdaguer~\cite{Dorca:1993si} studied the asymptotic
form of the trajectory originally discussed by Davies and
Fulling~\cite{Davies:1976hi,Davies:1977yv}.  Their procedure involved wave
packets for modes in both the \textit{in} and \textit{out} states.  This
technique allowed them to obtain a finite spectrum even though the total
number of particles produced by the mirror in that model is infinite.

Here we use wave packets only on the modes that approach future null
infinity to the right, $\mathscr{I}^{+}_R$, and are in the \textit{out}
vacuum state.  The idea is to model what a particle detector at a large
distance from the mirror would see.  It turns out that one can compute either
the packets first and then the Bogolubov transformation to determine the
number of particles in a given packet or one can do the computation in
reverse order.  We do the latter and first compute the exact Bogolubov
transformation for the modes and integrate over frequencies using the
appropriate weighting function~\cite{Fabbri:2005mw} to obtain the Bogolubov
coefficient for a packet.  To obtain the total number of particles in the
frequency range for that packet we then integrate over the entire frequency
range $\omega'$ of the modes in the \textit{in} vacuum state.  For two of
the trajectories considered there is an infrared divergence in the number
of particles created, which manifests as a divergence in the number of
particles in those packets that include modes of arbitrarily small
frequencies.  Any real
detector would have an infrared cutoff, so we eliminate this divergence
through the use of a low frequency cutoff in the computation of the wave
packets.  That is, for these trajectories, we do not consider packets that
include modes in the \textit{out} state with arbitrarily small frequencies.
In two other cases the Bogolubov transformation results in an infrared
divergence in the total number of particles created but there is no
corresponding divergence in the number of particles in the packets with
small frequency modes.  In the other two cases there are no infrared
divergences and the total number of particles is finite.

The outline of this paper is as follows.  In Section \ref{sec:model}, we
review the relevant aspects of the moving mirror model in (1+1) dimensions
in some detail.  In Section \ref{sec:twoprevious}, we discuss both the
Carlitz-Willey~\cite{Carlitz:1986nh} and Walker-Davies~\cite{Walker:1982}
trajectories.  For the Carlitz-Willey trajectory we first review some
previous results and provide an analytic expression for the trajectory.  We
then use wave packets to show that the spectrum is time independent with a
Planck character.  We analytically compute the spectrum at a given time for
wave packets of various frequency widths to investigate the distortion of
the spectrum.  For the Walker-Davies trajectory we review some of the
analytic results found previously.  Unlike the Carlitz-Willey case, it does 
not appear possible to calculate analytically the expected number of 
particles in the wave packet modes.  In Section \ref{sec:newmirrors}, we
discuss four new trajectories and for each compute the relevant Bogolubov
coefficients, the stress-energy tensor, the number of particles produced
at a given frequency, and for one trajectory, the total number of
particles produced.  Section \ref{sec:analysis} includes a comparison of
the time dependence of the stress-energy tensor for the quantum field and
the number of particles produced, where the latter is investigated through
the use of wave packets.  The difficulties encountered in obtaining
simultaneously good time and frequency resolution for the number of
particles produced are discussed, as well as the divergences that occur for
some trajectories in the number of particles produced and the energy of those
particles.  Our conclusions are given in Section \ref{sec:conclusions}.
Throughout this paper units are used such that $\hbar = c = 1$.

\section{\label{sec:model} BACKGROUND}

The moving mirror model in (1+1) dimensions consists of a quantized massless
scalar field in flat space that obeys Dirichlet boundary conditions on
a perfectly reflecting (mirror) boundary.  The scalar field satisfies the
wave equation
\be \Box \Phi = 0  \;. \ee
In this paper we always expand the field in terms of mode functions that
are parameterized by the frequency $\w$.  Denoting them for the moment as
$\phi_\omega$, they obey the equation
\be \left(- \partial_t^2 + \partial_x^2 \right) \phi_\w
= -\, \partial_u \partial_v \phi_\w = 0 \;, \ee
with
\bes
\bea u &\equiv& t - x \;, \\
     v &\equiv& t + x  \;.
\eea
\ese
The general solution is
\be
\phi_\w = g(v) + h(u) \;,
\ee
with $g$ and $h$ being arbitrary functions.  The mode functions are
normalized using the scalar product
\be
(\phi_1,\phi_2) \equiv - i \int_\Sigma \; d \Sigma \, n^\mu
\left[\phi_1(x) \stackrel{\leftrightarrow}{\partial_\mu}\phi^*_2(x)\right] \;.
\label{scalar-product}
\ee
Here $\Sigma$ is any Cauchy surface for the spacetime, $n^\mu$ is a
future-directed unit normal to that surface~\cite{Fabbri:2005mw}, and
we adopt the usual notation for the derivative acting to the right
first and then to the left with a minus sign~\cite{Birrell:1982ix}.  For
this scalar product the canonical relations hold,
\bes
\bea
(\phi_\omega(x),\phi_{\omega'}(x)) &=& - (\phi^{*}_\w(x),\phi^{*}_{\w'}(x))
= \delta(\w-\w')  \;, \\
(\phi_\w(x), \phi^{*}_{\w'}(x)) &=& 0 \;.
\eea
\label{normalization}
\ese

For Minkowski space with no boundaries we can choose the normalized modes
\bes
\bea
\phi_{\omega u} &=& \frac{1}{\sqrt{4 \pi \omega}}e^{- i \omega u},
\label{flat-u}  \\
\phi_{\omega v} &=& \frac{1}{\sqrt{4 \pi \omega}}e^{- i \omega v} \;.
\label{flat-v}
\label{flat-modes}
\eea
\ese
Then
\be
\Phi = \int_0^\infty \frac{d \omega}{\sqrt{4 \pi \omega}}
\left[ a_{\omega u} e^{- i \omega u}  + a_{\omega v} e^{- i \omega v} +
a^\dagger_{\omega u} e^{+ i \omega u} + a_{\omega v}^\dagger e^{+ i \omega v}
\right] \;,
\ee
with $a_{\omega u}$, $a_{\omega v}$, $a^\dagger_{\omega u}$, and
$a^\dagger_{\omega v}$ being the usual annihilation and creation operators.

If there is a mirror with trajectory $z(t)$, so that at any time $t$ the
mirror is at the position $x = z(t)$, then the spacetime effectively has a
boundary.  In this paper we only consider solutions to the mode equation
that are to the right of the mirror and incorporate reflection from the
mirror's surface.  We also only consider mirror trajectories that begin at
past timelike infinity, $i^-$.  In this case past null infinity,
$\mathscr{I}^-$, only consists of the surface $u = - \infty$.  This is a
Cauchy surface.  If the mirror trajectory ends at future timelike infinity,
$i^+$, then future null infinity, $\mathscr{I}^+$, only consists of the
surface $v =  \infty$ and this is a Cauchy surface.  But if the trajectory
is asymptotic to the null ray $v = v_0$, then $\mathscr{I}^+$ has two parts,
$\mathscr{I}^+_R$ and $\mathscr{I}^+_L$, using the notation of
Ref.~\cite{Carlitz:1986nh}.  The surface $\mathscr{I}^+_R$ lies at
$v = \infty$ and $\mathscr{I}^+_L$ consists of the part of the surface
$u = \infty$ which goes from $v = v_0$ to $v = \infty$.  Taken together they
also provide a Cauchy surface.

It is useful to evaluate the scalar product~\eqref{scalar-product} using the
Cauchy surfaces for $\mathscr{I}^-$ and $\mathscr{I}^+$.  It can be shown
(see e.g.~\cite{Good:2011}) that for $\mathscr{I}^-$
\be
(\phi_1,\phi_2) = -i \int_{-\infty}^\infty \left[\phi_1(u=-\infty,v)
\stackrel{\leftrightarrow}{\partial_v}
\phi^*_2(u= -\infty,v) \right] dv  \;,
\label{scalar-v}
\ee
and for $\mathscr{I}^+$
\bea
(\phi_1,\phi_2) &=& -i \int_{-\infty}^\infty \left[\phi_1(u,v=\infty)
\stackrel{\leftrightarrow}{\partial_u} \phi^*_2(u,v=\infty) \right] du
\nonumber \\
& &  -i \int_{v_0}^\infty \left[\phi_1(u=\infty,v)
\stackrel{\leftrightarrow}{\partial_v} \phi^*_2(u=\infty ,v) \right] dv \;.
\label{scalar-u}
\eea

If Dirichlet boundary conditions are imposed on the scalar field, then the
mode functions $\phi_\w$ must vanish at the location of the mirror.  To
quantify this, it is useful to introduce functions $u_m(t)$ and $v_m(t)$
that give the values of $u$ and $v$ at the location of the mirror at a
given time $t$.  Thus
\bes
\bea
u &=& u_m(t) = t - z(t) \;,  \label{um} \\
v &=& v_m(t) = t + z(t) \;.  \label{vm}
\eea
\ese
We can invert the first equation to get $t$ via a function $t_m(u)$ or we
can invert the second to get $t$ via a function $\bar{t}_m(v)$.

In the presence of a mirror we can consider either the mode functions that
are positive frequency at $\mathscr{I}^-$, and thus correspond to the
\textit{in} vacuum state, or the mode functions that are positive frequency
at $\mathscr{I}^+$, and thus correspond to the \textit{out} vacuum state.
The modes that are positive frequency at $\mathscr{I}^-$ are
\be
\phi^{\rm in}_{\omega'} = \frac{1}{\sqrt{4 \pi \omega'}}
\left[ e^{- i \omega' v} - e^{- i \omega' p(u)} \right] \;.
\label{p-modes}
\ee
Substitution into Eq.~\eqref{normalization} using Eq.~\eqref{scalar-v}
shows that the normalization is correct.  For these mode functions to
vanish at the mirror we must have $v = p(u)$ \emph{at the location of the
mirror}.  If we invert Eq.~\eqref{um} above to find $t = t_m(u)$ and
then use the definition~\eqref{vm}, we find that
\be
p(u) = t_m(u) + z(t_m(u)) ,
\label{pumeq}
\ee
which fixes the function $p(u)$.

In a general left-right construction (for mirrors that have a horizon at
$v_0$), there are two sets of mode functions that are positive frequency
at $\mathscr{I}^+$.  One set, which we will denote as $\phi^R_\omega$, are
nonzero at $\mathscr{I}^+_R$ and zero at $\mathscr{I}^+_L$.\footnote{The
left and right coefficient formulation lies at the crux of the calculations
in Ref.~\cite{Carlitz:1986nh} and despite the call for more
attention~\cite{Wilczek:1993jn}, the construction has been under-utilized.}
The other set, $\phi^L_\omega$, are zero at $\mathscr{I}^+_R$ and nonzero
at $\mathscr{I}^+_L$.  The former are given by
\be
\phi_\omega^{\rm R,\;out} = \frac{1}{\sqrt{4 \pi \omega}} \,
\left[ e^{-i \omega f(v)} - e^{-i \omega u} \right] \;, \qquad v < v_0 \;.
\label{f-modes}
\ee
Mirrors that are asymptotically inertial in the future have $v_0 = \infty$.
Note that all trajectories that begin at past timelike infinity, $i^-$,
(the only type we consider here) span the range
$-\infty < u < \infty$.  Substitution into Eq.~\eqref{normalization} using
Eq.~\eqref{scalar-u} shows that these modes are normalized correctly also.
Again, for these modes to vanish at the mirror, we must have
$u = f(v)$ \emph{at the location of the mirror}
which is the inverse relation to $v = p(u)$ and an equivalent
representation of the mirror trajectory in $u$, $v$ coordinates.  If
Eq.~\eqref{vm} is inverted to find $t = \bar{t}_m(v)$ and the
definition~\eqref{um} is used, then one finds that
\be
f(v) = \bar{t}_m(v) - z(\bar{t}_m(v))
\label{fvmeq} \;,
\ee
which fixes the function $f(v)$.

There are no other modes if $v_0 = \infty$.  But if the mirror's trajectory
is asymptotic to the null surface $v = v_0$, then one must also include the
set of modes $\phi^{L}_\omega$ that reach $\mathscr{I}^+_L$ and never
interact with the mirror.  Substituting into Eq.~\eqref{normalization} and
using Eq.~\eqref{scalar-u} one finds that
\be
(\phi^L_\omega,\phi^L_{\omega'}) = -i \int_{v_0}^\infty dv \phi^L_\omega
\stackrel{\leftrightarrow}{\partial_v} \phi^{L*}_{\w'} = \delta(\w-\w') \;.
\label{phi-L-norm}
\ee
To further examine the behavior of these modes it is useful to work with a
specific trajectory.  This has been done by Carlitz and
Willey~\cite{Carlitz:1986nh} for a trajectory with a future horizon. We study
other aspects of this trajectory in Sec.~\ref{sec:twoprevious}.

The usual procedure for calculating interesting observable quantities, such
as the energy or particle number, starts with the choice of an appropriate
trajectory, $z(t)$.  Then either the function $p(u)$ or its inverse $f(v)$
is found using the procedures described above.  A key aspect of these
procedures involves the solution of the relevant, and sometimes
transcendental, function inversions.  This requirement has made it difficult
to find trajectories that allow both a fully analytic description of the
mirror's motion and an analytic calculation of the associated Bogolubov
coefficients (see below).

The function $p(u)$, commonly called the ray-tracing
function~\cite{Reuter:1988qh}, characterizes the mirror trajectory and is
incorporated in the modes, the two-point function, the energy flux, and the
correlation functions.  The trajectories and, where known, the ray-tracing
functions for the mirrors considered in this paper are given in
Table \ref{restab}.

\begin{table}[t]
\setlength{\extrarowheight}{3pt}
\begin{center}
\caption{Some classes of trajectories, $z(t)$, and the ray-tracing functions
$p(u)$ and $f(v)$ associated with them.}
\label{restab}
\begin{tabularx}{\textwidth}{XXXX}
\hline
\hline
 & Trajectory  &  $p(u)$ &  $f(v)$    \\
\noalign{\smallskip}
\hline
Static     & $z = 0$ &  $ p = u$ & $ f = v$ \\
Constant velocity    & $z =  -v_0 t$ &  $ p = \f{1-v_0}{1+v_0} u$
& $ f = \f{1+v_0}{1-v_0}v $  \\
Uniform acceleration & $ z =  \kp^{-1}-\sqrt{\kp^{-2}+t^2}$
&  $ p = \f{u}{1+\kappa u}$ & $f = \f{ v }{1-\kappa v}$ \\
Carlitz-Willey      & $z =  -t - \f{1}{\kp}W(e^{-2\kappa t})$
&  $p =  -\f{1}{\kp}e^{-\kappa u}$ & $f = - \frac{1}{\kappa} \log(-\kappa v)$ \\
Walker-Davies & $t = -z \pm A \sqrt{e^{-2z/B} - 1} $ & & \\
Arctx     & $ z =  -\f{1}{\mu} \tan^{-1} ( e^{\mu t })$ &   & \\
Darcx     & $ z = -\f{\xi}{\nu} \sinh^{-1} ( e^{\nu t })$ &   &  \\
Proex     & $ z = -\f{1}{\rho}W(e^{\rho t})$
& $ p = u-\f{1}{\rho}W(2e^{\rho u})$ & $f = v + \frac{2}{\rho} e^{\rho v} $  \\
Modified Carlitz-Willey      & $z = -\frac{1-\sigma}{1 + \sigma} t $
&  $ p = \sigma u -\f{1}{\kp}e^{-\kappa u}$ &
$ f = \frac{v}{\sigma} + \frac{1}{\kappa} W(e^{-v \kappa/\sigma}/\sigma)$  \\
  & \hspace{0.5cm} $  - W[e^{-2\kappa t}/(1 + \sigma)]/\kappa $  &  &  \\
\noalign{\smallskip}
\hline
\hline
\end{tabularx}
\end{center}
\end{table}

\subsection{Bogolubov transformations}

One way to describe the particle production that arises in the presence of
an accelerating mirror is to use the Bogolubov transformation.  The positive
frequency modes at $\mathscr{I}^{-}$, $\phi^{\rm in}_{\w'}$, form a complete
set and one can expand modes at $\mathscr{I}^{+}$ in terms of them,
\be
\phi^J_\omega = \int_0^\infty d \omega' \left[ \alpha^J_{\omega \omega'}
\phi^{\rm in}_{\omega'} + \beta^J_{\omega \omega'}
\phi^{\rm in \;*}_{\omega'} \right] \;,
\ee
with $J$ representing either $R$ or $L$.  Using the
relations~\eqref{normalization} one finds
\bes
\bea
\alpha^J_{\omega \omega'} &=& (\phi^J_\omega,\phi^{\rm in}_{\omega'}) \;,
\label{alpha-def} \\
\beta^J_{\omega \omega'} &=& -(\phi^J_\omega,\phi^{{\rm in}\; *}_{\omega'})
\label{beta-def} \;.
\eea
\label{alpha-beta-def}
\ese
The field $\Phi$ expressed in terms of the mode functions can be represented
in either of two ways,
\bes
\bea
\Phi &=& \int_0^\infty d \omega' \left[a^{\rm in}_{\w'}
\phi^{\rm in}_{\w'} +  a^{\rm in \; \dagger}_{\w'}
\phi^{\rm in \; *}_{\w'} \right] \\
     &=&  \sum_J  \int_0^\infty d \omega \left[b^J_\w \phi^J_\w
+  b^{J \; \dagger}_\w \phi^{J \; *}_\w \right] \;.
\eea
\ese
Using $b^J_\w = (\Phi,\phi^J_\w)$~\cite{Fabbri:2005mw} one finds
\be
b^J_\w = \int_0^\infty d \w'
\left[ (\alpha^J_{\w \w'})^* a^{\rm in}_{\w'}
- (\beta^J_{\w \w'})^* a^{{\rm in} \; \dagger}_{\w'} \right]  \;.
\ee
If the field is in the \textit{in} vacuum state specified by the positive
frequency modes at $\mathscr{I}^-$, we can use the operator
$N^J_\w \equiv (b^J_\w)^\dagger  b^J_\w$ to compute the average number of
particles with frequency $\w$ that reach $\mathscr{I}^{+}_J$,
\be
\langle N^J_\w \rangle \equiv \langle 0_{\rm in} | N^J_\w | 0_{\rm in} \rangle
= \int_0^\infty d \w' \, | \beta^J_{\w \w'} |^2 \label{occ} \;.
\ee
The expectation value of the total number of particles that reach $\mathscr{I}^{+}_J$ is
\be
\langle N^J \rangle \equiv \langle 0_{\rm in} | N^J | 0_{\rm in} \rangle
= \int_0^\infty d \w \, \int_0^\infty d \w' \, | \beta^J_{\w \w'} |^2  \;.
\label{Ntotal}
\ee

Since we are primarily concerned with the number of particles that reach
$\mathscr{I}^{+}_R$, we will focus on the computation of $\beta^R_{\w \w'}$.
If the Cauchy surface $\mathscr{I}^-$ is used, then Eq.~\eqref{beta-def}
along with Eqs.~\eqref{scalar-v},~\eqref{p-modes}, and~\eqref{f-modes} gives
\be
\label{betaf}
\beta^R_{\w\w'} = \f{1}{4\pi\sqrt{\w\w'}}\int_{-\infty}^{v_0} dv \;
e^{-i\w' v -i\w f(v)} \left(\w' - \w \frac{d f(v)}{dv} \right) \;.
\ee
If the Cauchy surface $\mathscr{I}^{+}_R$ is used, then one similarly finds
\be
\label{betap}
\beta^R_{\w\w'} = \f{1}{4\pi\sqrt{\w\w'}}\int_{-\infty}^{\infty} du \;
e^{-i\w  u - i\w'p(u)} \left(\w' \frac{ d p(u)}{du} - \w \right) \;.
\ee
These expressions are of course equivalent.

It is possible to write Eq.~\eqref{betap} in terms of a
time integral over a function of the trajectory $z(t)$.  Since $p(u)$ is a
fixed function of $u$ and $u$ ranges from $-\infty$ to $+\infty$, one can
substitute $u_m(t)$ for $u$ in Eq.~\eqref{betap}.  Then using Eq.~\eqref{um}
to change variables, one finds
\be
\label{betaz}
\beta^R_{\w\w'} =
\f{1}{4\pi\sqrt{\w\w'}}\int_{-\infty}^{\infty} dt \;
e^{-i\w_{+}t + i\w_{-} z(t)} \left(\w_{+}\dot{z}(t)-\w_{-}\right) \;,
\ee
where $\w_+ \equiv \w + \w'$ and $\w_{-} \equiv \w-\w'$.  Note that if we
consider an inertial trajectory it is easy to show that
$\beta^R_{\w\w'} = 0$.  Thus, as expected no particles are produced when
the mirror does not accelerate.

It is not hard to show that the Bogolubov coefficient $\alpha_{\w \w'}$ may
be obtained from the above expressions for $\beta_{\w \w'}$ by letting
$\w' \rightarrow -\w'$ everywhere in the expressions for $\beta_{\w \w'}$,
except for the factor $1/\sqrt{\w\w'}$ which must remain unchanged.  Finally,
we note that if the trajectory is initially inertial and the acceleration
does not continue forever, then the total energy produced is finite and
given by the following sum over the quantum modes~\cite{Walker:1984vj}
\be
E_{qs} = \int_{0}^{\infty} \w \la N_{\w} \ra d\w \;.
\label{Eqs}
\ee

\subsection{Stress-energy tensor}

The renormalized stress-energy tensor for the massless, minimally coupled
scalar field was
computed in terms of the function $p(u)$ and in terms of the trajectory
$z(t)$ by Davies and Fulling~\cite{Davies:1976hi}.  They found that the
energy flux produced by the mirror as a function of $u$ is given
by\footnote{Some other components are
$\la T^{tt} \ra = \la T^{tx} \ra = \la T^{x x}\ra = \la T_{uu} \ra$.}
\be
\label{stress}
\la T_{uu} \ra =
\f{1}{24\pi}\left[\f{3}{2}\left(\f{p''}{p'}\right)^2-\f{p'''}{p'}\right] \;,
\ee
where primes indicate derivatives with respect to $u$.  Their expression for
the energy flux in terms of $z(t_m(u))$ is
\be
\label{stressz}
\la T_{uu} \ra =
\left.\f{\dddot{z}(\dot{z}^2-1)-3\dot{z}\ddot{z}^2}
{12\pi(\dot{z}-1)^4(\dot{z}+1)^2}\right|_{t=t_m(u)} \;,
\ee
where the dots refer to derivatives with respect to $t$.  Eq.~\eqn{stressz}
is equivalent to Eq.~\eqn{stress} evaluated at the surface of the mirror.
In either case, it is easy to show that for an inertial trajectory
$\la T_{uu} \ra= 0$, as would be expected when the scalar field is in the
vacuum state.

It is also possible to write Eq.~\eqref{stressz} in terms of the time
derivative of the proper acceleration\footnote{The proper acceleration is
the acceleration in the instantaneous rest frame of the mirror.  Note that
the time derivative is in the inertial frame, not the rest frame of the
mirror.}
\be
\alpha \equiv \ddot{z}/(1- \dot{z}^2)^{3/2} \;.
\label{accel-def}
\ee
The result is
\be
\la T_{uu} \ra = -\f{\dot{\alpha}}{12\pi} \, \f
{\sqrt{1+\dot{z}}}{(1-\dot{z})^{3/2}} \;,
\label{stress-alpha}
\ee
The overall negative sign implies that on the right hand side of the mirror
a flux of negative energy is given off if the change in acceleration of the
mirror is towards the right and a flux of positive energy is given off if
the change in acceleration is towards the left.

For trajectories that are asymptotically inertial in the limits
$t \rightarrow \pm \infty$, $p(u)\rightarrow c_1 u + c_0$ for some constants
$c_1$ and $c_0$.  For trajectories considered in this paper, a finite amount of energy reaches
$\mathscr{I}^+_R$ (except for the Carlitz-Willey or modified Carlitz-Willey classes).  Because the flux~\eqref{stress} is only a function of
$u$ and therefore does not fall off at $\mathscr{I}^+_R$, the total amount
of energy $E_{st}$ that reaches $\mathscr{I}^+_R$ can be obtained by
integrating over $u$,
\be
\label{Efromp}
E_{st}  = \int_{-\infty}^{\infty} \la T_{uu} \ra \; du  \;.
\ee
Walker~\cite{Walker:1984vj} has a proof that $E_{st} = E_{qs}$ provided that the mirror is asymptotically inertial in both the past and future, i.\ e.\ $\alpha(\pm\infty)= 0$, and the velocity \textit{towards} $\mathscr{I}^+_R$ never reaches the speed of light, $\dot{z}(\pm\infty)\neq 1$.  Substituting Eq.~\eqref{stress} into~\eqref{Efromp} and integrating by parts one finds
\be
E_{st}=\f{1}{48\pi}\int_{-\infty}^{\infty}\left(\f{p\,''}{p\,'}\right)^2 du \;,
\label{E-p}
\ee
so long as the surface terms with ${p\,''}/{p\,'}$ vanish as
$u\rightarrow \pm\infty$.  The result can also be written in terms of a
time integral over a function of the trajectory of the mirror by letting
$u \rightarrow u_m(t)$, using Eq.~\eqref{um} to invert to find
$t = t_m(u)$, and then using Eq.~\eqref{stressz}.  The result is
\be
\label{Efromj}
E_{st} =
-\f{1}{12\pi}\int_{-\infty}^{\infty}\dot{\alpha}
\sqrt{\f{1+\dot{z}}{ 1-\dot{z}}} \;dt \;.
\ee
Integrating by parts gives another expression,
\be
\label{Epartsz}
E_{st} = \f{1}{12\pi}\int_{-\infty}^{\infty} \alpha^2 (1+ \dot{z}) dt =
\f{1}{12\pi}\int_{-\infty}^{\infty}
\f{\ddot{z}^2}{(1+\dot{z})^2(1-\dot{z})^3} \;dt \;.
\ee
Notice that this last form masks the dependence on the time derivative of
the proper acceleration.

\subsection{Wave packets}

Another way to investigate particle production is to use wave
packets~\cite{Hawking:1974sw, Fabbri:2005mw}.  An advantage of this
approach, as discussed in the Introduction, is that one can study the
time-dependent aspects of particle production.

A wave packet, $\phi_{jn}$, can be constructed from $\phi_\w$ by integrating
over a finite range of frequencies with a particular weighting function so
that~\cite{Fabbri:2005mw}
\be
\phi_{jn} \equiv \f{1}{\sqrt{\epsilon}}\int_{j\epsilon}^{(j+1)\epsilon} d\w\;
e^{2\pi i \w n/\epsilon} \phi_\w \;.
\label{mode-packet}
\ee
Here $n$ takes on integer values and $j$ takes on nonnegative integer values.
Substituting Eq.~\eqref{f-modes} into Eq.~\eqref{mode-packet} and noting
that the first term does not contribute due to rapid oscillations in the
limit $v \rightarrow \infty$, one can see that the integral is largest for
values close to $ u = 2 \pi n/\epsilon$.  It is clear
from~\eqref{mode-packet} that the value of $j$ is related to the frequency
of the modes in the packet with $(j+1/2)\epsilon$ giving the frequency at
the center of the range and $\epsilon$ giving the width of the range.  When
the weighting functions are applied to the modes
$e^{-i \omega u}/\sqrt{4 \pi \w}$, the
resulting wave packets form a complete and orthonormal set.

One can use the scalar product to construct the Bogolubov coefficients that
correspond to the wave packets~\cite{Fabbri:2005mw}.  As mentioned
previously, we are concerned in this paper with the particles that reach
$\mathscr{I}^+_R$.  In that case
\be
\beta^R_{j n, \w'} = - (\phi^R_{j \, n}, \phi^{\rm in \; *}_{\w'}) \;.
\ee
It is possible to obtain these wave packet coefficients directly from the
coefficients $\beta^R_{\w\w'}$ by using the same weighting, integrating
over frequency, and swapping the order of integration
\be
\beta^{R}_{j n, \w'} =
\f{1}{\sqrt{\epsilon}}\int_{j\epsilon}^{(j+1)\epsilon}
d\w \; e^{2\pi i \w n/\epsilon} \beta^{R}_{\w\w'} \;.
\label{beta-packet}
\ee
The average number
of particles produced for given values of $n$ and $j$ is
\bea
\la N^R_{jn}\ra &=& \int_0^\infty d\w' |\beta^{R}_{jn,\w'}|^2
\nonumber
\\
&=& \int_0^\infty d\w' \int_{j \epsilon}^{(j+1)\epsilon}
\frac{d \w_1}{\sqrt{\epsilon}} \int_{j \epsilon}^{(j+1)\epsilon}
\frac{d \w_2}{\sqrt{\epsilon}} e^{2 \pi i(\w_1- \w_2)n/\epsilon}
\beta^R_{\w_1 \w'} \beta^{R \; *}_{\w_2 \, \w'}  \;.
\label{Njn}
\eea
This quantity gives the average number of particles that reach
$\mathscr{I}^+_R$ in the frequency range
$j \epsilon \le \omega \le (j+1) \epsilon$ and in the approximate time
range $ (2\pi n - \pi)/\epsilon \le u \le (2 \pi n + \pi)/\epsilon$.  It can
be used to estimate the average number of particles that a detector would
see in this frequency range if it was turned on during the above time
period near an event centered at some $x$ and some large $v$.  Thus
computation of $\la N_{jn} \ra$ for a range of values of $j$ and $n$
allows one to construct the evolution of the spectrum of the produced
particles in time, to the extent allowed by the uncertainty relation, as it
would be seen by a series of particle detectors spread out over a line of
constant but large $v$.

Note that one can also estimate the total energy of the particles produced
by multiplying the number of particles in a given bin by the frequency at
the center of that bin,
\be
E_{ep} = \sum_{j,n}
\left( j +\frac{1}{2}\right)\epsilon \; \la N_{jn}\ra \;. \label{Eep}
\ee
This estimate of the energy can be compared to the energy of
particles produced $E_{qs}$ or the total stress-energy flux $E_{st}$ to test
the accuracy of the wave packet description of particle production.

\section{Two Previously-Studied Mirror Trajectories}
\label{sec:twoprevious}

In this section we examine two previously-studied types of mirror
trajectories.  One of these, developed by Carlitz and
Willey~\cite{Carlitz:1986nh}, consists of a trajectory
that has a future horizon at $v = v_0 = 0$ but no past horizon.  The
trajectory is designed to yield a constant stress-energy flux.  The
functional form of the trajectory allows many quantities of interest to be
computed analytically.  The second type is the class of trajectories studied
by Walker and Davies~\cite{Walker:1982}.  For these trajectories, the mirror
begins and ends asymptotically at rest.  Thus the total number of particles
produced is finite.  In this case too, a number of quantities can be obtained
analytically.  For both of these types of trajectories we have extended
the analysis by using wave packets to compute the spectrum of created
particles.

\subsection{Carlitz-Willey trajectory}
\label{sec:cw}

In their paper~\cite{Carlitz:1986nh} Carlitz and Willey point out that if
the motion of the mirror is specified (in $u$ and $v = p(u)$ coordinates)
by taking the ray-tracing function to be
\be
p(u) = - \f{1}{\kappa} e^{- \kappa u} \;,
\label{p-carlitz-willey}
\ee
then a constant energy flux results.  Substitution into Eq.~\eqref{stress}
gives the energy flux in terms of the free parameter $\kappa$,
\be
\la T_{uu} \ra = \frac{\kappa^2}{48 \pi}  \;.
\ee
An implicit functional form of the trajectory in $t$ and $x = z(t)$
coordinates can be obtained by substituting Eq.~\eqref{p-carlitz-willey}
into Eq.~\eqref{pumeq},
\be
t + z(t) = - \frac{1}{\kappa} e^{-\kappa t  + \kappa z(t)}  \;.
\ee
Carlitz and Willey did not provide the explicit functional form for $z(t)$.
However, we find that it can be given as
\be
z(t) = -t - \f{1}{\kp}W(e^{-2\kappa t})  \; ,
\label{cw-trajectory}
\ee
which involves the Lambert $W$ function (also known as the product logarithm).
A plot of this trajectory is given in Ref.~\cite{Carlitz:1986nh} and is
shown also in our Fig.~\ref{fig:trajects}.  It is not difficult to show
that $\dot{z}\rightarrow \pm 1$ in the limits $t \rightarrow \mp \infty$
and that $z < 0$ for all time.  The mirror trajectory begins at past timelike
infinity, $i^{-}$, and at late times approaches $v  = 0$.  Substitution into
Eq.~\eqref{accel-def} gives the proper acceleration,
\be
\label{thermalacc}
\alpha(t)= -\f{\kp}{2\sqrt{W(e^{-2 \kp t})}} \;,
\ee
which is not constant, even though
the energy flux is.

In Ref.~\cite{Carlitz:1986nh} analytic expressions were found for the
Bogolubov coefficients.  In particular it was found that\footnote{Here we
have adapted the expression given in~\cite{Carlitz:1986nh} to the
conventions we are using.}
\be
\beta^R_{\w\w'} =
\f{1}{4\pi\sqrt{\w\w'}}
\left\{-\f{2\w}{\kp}e^{-\pi\w/2\kp}
\left(\f{\w'}{\kp}\right)^{-i\w/\kp}
\Gamma\left[\f{i\w}{\kp}\right]\right\}
\label{beta-c-w} \;.
\ee
Thus
\be
\left| \beta^R_{\w\w'} \right|^2
= \f{1}{2\pi\kp \w'} \f{1}{e^{2\pi \w/\kp} - 1} \;,
\ee
resulting in both infrared and ultraviolet divergences for the quantity
$\la N^R_\w \ra$ in Eq.~\eqref{occ}.  This is not surprising given the fact
that the acceleration of the mirror, while zero in the limit
$t \rightarrow -\infty$, is nonzero at any finite time in the past.  The
radiation is produced with a thermal spectrum and propagates to
$\mathscr{I}^+_R$~\cite{Carlitz:1986nh}.\footnote{This thermal spectrum is
of course that of a one-dimensional black body.  A similar physical
manifestation of one-dimensional thermal radiation is the resistor.
Discovered by Nyquist \cite{Nyquist:1928zz} in 1928, a resistor in a
lossless transmission line of great length in equilibrium at temperature $T$
has thermal electric noise that is an analog of a black body in one
dimension.}

It is possible to compute analytically the expectation value of the wave
packet number
$\la N_{jn} \ra$.  First substitute Eq.~\eqref{beta-c-w} into
Eq.~\eqref{Njn}, interchange the order of integration, and make the variable
transformation $y = \ln \w'$.  Then integrating over $y$ gives a result that
is proportional to $\delta(\w_1- \w_2)$.  Next integrating over $\w_2$ and
setting $\w_1 = \w$ gives
\bea
\la N_{jn}\ra  &=& \frac{1}{\epsilon}\int_{j
\epsilon}^{(j+1)\epsilon} d\w \,\frac{1}{e^{2 \pi \w/\kappa}-1}
= \f{\kappa}{2\pi\epsilon} \ln \left(\f{e^{\f{2\pi(j+1)\epsilon}
{\kappa}}-1}{e^{\f{2\pi j\epsilon}{\kappa}}-1}\right) - 1  \;.
\label{Njn-c-w}
\eea
There is a divergence in the particle count in the lowest frequency bin,
$j = 0$.  This result is similar to the infrared divergence found previously
in $\la N^R_\w \ra$ and the reason for it is the same.  Physically of course,
one cannot measure particles of infinite wavelength.  Thus in a real
particle detector the lowest frequency bin would have a lower limit cutoff
rather than extending all the way to $\w = 0$.

It is also evident that for this mirror there is no dependence in
$\la N_{jn}\ra $ on the parameter $n$ and accordingly the average spectrum
of particles recorded by a detector would be independent of time.  This is
almost certainly related to the fact that for this trajectory the energy
flux is constant.

To find effects of various frequency ranges $\epsilon$ on the spectrum and
to recover the Planck form for the spectrum in the limit that
$\epsilon \rightarrow 0$, one can first write the expression
in~\eqref{Njn-c-w} in terms of $\w_j = (j + 1/2) \epsilon$ and then expand
it in powers of $\epsilon$ with the quantity $\w_j$ fixed.  The result is
\be
\label{planckian}
\la N_{jn}\ra = \f{1}{e^{2\pi\w_j/\kp}- 1}
\left[ 1 +\frac{ \pi^2 e^{2\pi\w_j/\kp}
(1 + e^{2\pi\w_j/\kp})}{6 \kappa^2 (e^{2\pi\w_j/\kp} - 1)^2}
\epsilon^2 + O(\epsilon^4) \right]
\;.
\ee
With $\epsilon$ fixed, the second term approaches
$\epsilon^2/(12 \omega_j^2)$ for small $\omega_j$ and behaves like
$\pi^2 \epsilon^2/(6 \kappa^2)$ for large values of $\omega_j$.  Thus, for
a given frequency width the deviation of the spectrum from the Planck form
becomes more pronounced for smaller values of the central frequency
$\omega_j$.

\subsection{Walker-Davies trajectory}

\begin{figure}
\includegraphics[scale=1.0]{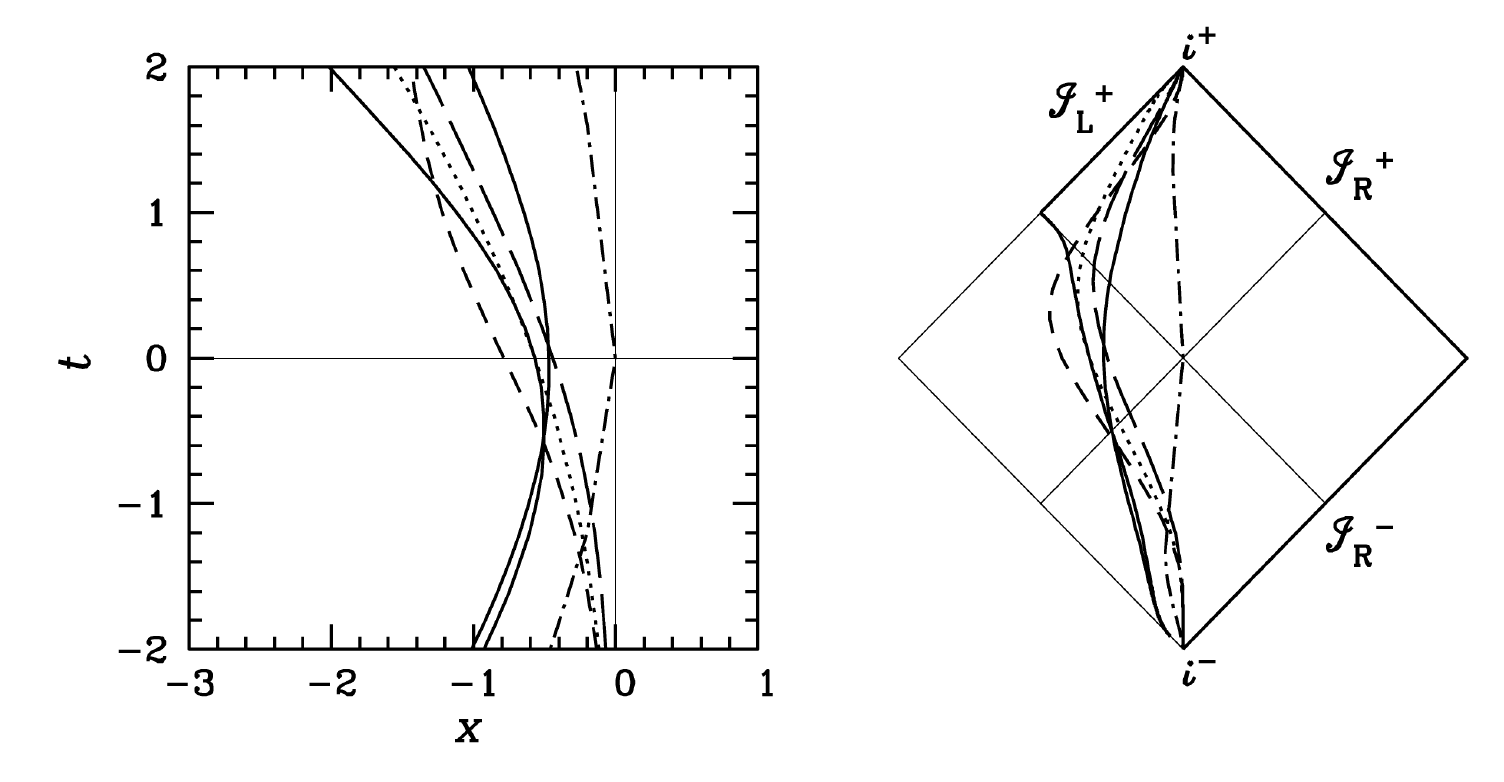}
\caption{\label{fig:trajects}
Six analytically-known mirror trajectories.  In the left panel the mirror
trajectories are plotted in $t$ and $x$ coordinates.  The right panel
depicts the same trajectories in a Penrose diagram.  The Carlitz-Willey
trajectory, with $\kappa = 1$, and the modified Carlitz-Willey trajectory,
with $\kappa = 1$ and $\sigma = 1/3$, are shown as solid curves.  The
Carlitz-Willey trajectory gives rise to a horizon at $v = 0$, while the
modified Carlitz-Willey trajectory ends at future timelike infinity
($i^+$).  The short-dashed curve indicates the Arctx mirror with $\mu = 1$,
which is static in the distant past and future.  The long-dashed curve is
the Darcx trajectory with $\nu = 1$ and asymptotic future velocity
$\xi = -1/2$.  The Walker-Davies trajectory for $A=2$, $B=1$ is shown as
the dot-dashed curve.  The Proex trajectory with $\rho = 1$ is denoted by
the dotted curve.  We consider only the region of spacetime to the right of
a mirror.  Thus past null infinity on the left (not labeled) plays no role
in our analysis.  Similarly, there is an abbreviated portion of future
null infinity on the left (${\mathscr{I}_{L}}^{+}$), but only in the case
of the Carlitz-Willey trajectory.  All of the other mirror trajectories
begin and end at $i^{-}$ and $i^{+}$, respectively. }
\end{figure}

The Walker-Davies trajectory~\cite{Walker:1982} is given by the relation
\be
t = -z \pm A \sqrt{e^{-2z/B} - 1} \;,
\ee
where $A>B$ and where the plus sign is adopted for $t>0$ and the minus sign
for $t<0$.  The trajectory is plotted in Fig.~\ref{fig:trajects}.  Note that
the curve is $C^\infty$ in spite of the change in sign across branches for
positive and negative values of $t$.  The mirror begins at $i^-$ at rest with
$t,z = - \infty$.  It accelerates to the right and then decelerates back to
rest at $t = z = 0$.  Then it first accelerates and then decelerates to the
left, ending at rest at $i^+$ at $t = \infty$ and $z = -\infty$.  The modes
to the right of the mirror always end at $\mathscr{I}^+_R$.  Therefore
we drop the subscript $R$ in what follows.

Walker and Davies~\cite{Walker:1982} calculated the stress tensor and found
that \footnote{Note the misprint in their expression for the numerator.}
\be
\la T_{uu} \ra = \frac{B}{6 \pi} \frac{\left(\chi^5
+ \frac{1}{2} B \chi^4 - 2 A^2 \chi^3 - 3 B A^2 \chi^2 - 3 A^4 \chi
- \frac{3}{2} A^4 B\right)}{(\chi^2 + 2 B \chi + A^2)^4} \;,
\label{Tuu-WD}
\ee
where $\chi$ is a parameter related to the null coordinate $u$ by
\be
u = B \ln \left( \chi^2/A^2 + 1 \right) + \chi \; .
\ee
By integrating the flux over all $u$, they found the total energy
to be
\be
E_{st} = \f{B^2}{48(A^2-B^2)^{3/2}}\;.
\label{Estwalker}
\ee
They also were able to calculate the Bogolubov coefficients and found that
\be
\label{walkb} \left| \beta_{\w\w'} \right|^2
= \f{2AB}{\pi^2}\left(\f{\w'}{\w'+\w}\right)\sinh(\pi \w B)|K_{q}(r)|^2 \;,
\ee
where $q\equiv -\f{1}{2}+i\w B$, $r\equiv A(\w'+\w)$, and $K_{q}(r)$ is a
modified Bessel function of the second kind.  We have not found an analytic
expression for $\la N_\omega \ra$ in Eq.~\eqref{occ} and therefore also do
not have one for $E_{qs}$.  Furthermore, we do not have an analytic expression
for $\la N_{jn} \ra$ in this case and instead have computed it numerically.
We defer until Sec.~\ref{sec:analysis} discussion of these results.

\section{New Trajectories}
\label{sec:newmirrors}

In this section, we consider four new types of trajectories for which it is
possible to analytically compute the Bogolubov coefficients
$\beta_{\w \w'}$.  Having this analytic result in turn makes it feasible to
compute numerically the expectation value of the particle number
$\la N_{jn} \ra$ associated with the wave packets.  Trajectories of each
type are plotted in Fig.~\ref{fig:trajects} for specific parameter choices.
The functional form of the trajectories and their corresponding ray-tracing
functions, where known, are summarized in Table \ref{restab}.

\subsection{Arctx mirror trajectory}

As seen in the previous section the Walker and Davies~\cite{Walker:1982}
class is composed of trajectories that begin and end at rest in the limits
$t \rightarrow \pm\infty$.  They produce a finite number of particles and
a finite amount of energy.  Another trajectory with these features can be
devised by taking
\be
z(t) = -\f{1}{\mu}\tan^{-1}(e^{\mu t})  \;.
\label{arctx-z}
\ee
Here $\mu$ is a positive constant.  Such a mirror starts at rest at
$x = 0$ in the infinite past and ends at rest at $x = -\pi /2 \mu$ in the
infinite future.  We refer to this trajectory as Arctx, drawn from
\textbf{Arct}angent E\textbf{x}ponential.

The functional form of this trajectory is simple enough that a number of
properties can be derived analytically.  For example the proper acceleration
is given by
\be
\alpha(t) =
\mu \f{ 4\; \text{sech}(\mu t) \tanh(\mu t)}
{\left[4-\text{sech}(\mu t)\right]^{3/2}} \;.
\label{arctx-accel}
\ee
Starting from zero in the $t\rightarrow -\infty$ limit, the proper
acceleration is negative for $t <0$.  It first increases and then decreases
in magnitude, before reaching zero at $t = 0$.  For positive values of $t$
it is positive, and again first increases and then decreases to zero in
the limit $t \rightarrow \infty$.  A second important quantity, the
stress-energy, can also be obtained.  Substitution of~\eqref{arctx-z}
into~\eqref{stressz} gives
\be
\label{tuTuu}
\la T_{uu} \ra =
\frac{\mu^2\cosh[\mu t_m(u)](-5-2 \cosh[2\mu t_m(u)]+\cosh[4\mu t_m(u)])}
{3 \pi (1-2 \cosh[\mu t_m(u)])^2 (1+2 \cosh[\mu t_m(u)])^4} \; .
\ee
The finite total energy would follow from integrating this flux
over $u$, but it is simpler to substitute~\eqref{arctx-z} into~\eqref{Epartsz}
to find
\be
E_{st} = \f{\mu}{2592\pi} (13 \sqrt{3} \pi -36) \;.
\label{E-Tuu-arctx}
\ee
Next, it proves possible to calculate analytically the Bogolubov
coefficients.  Substituting~\eqref{arctx-z} into Eq.~\eqref{betaz} we find
\be
\beta_{\w\w'} =
g_0 \left[g_1 \Gamma(-m) \Gamma(-q)-g_2 \Gamma(m) \Gamma(q)\right] \; ,
\label{beta-arctx}
\ee
where
\bes
\bea
q &\equiv& \f{i}{\mu}(\w' + \w) \;, \\
m &\equiv& \f{1}{2\mu}(\w'-\w) \;, \\
g_0 &\equiv& i e^{ \frac{i \pi q}{2} } \frac{\sqrt{ 4 m^2+q^2}}{2 \pi \mu}
\f{\text{sin}(\pi m)}{\text{sin}[\pi(m+q)]} \; , \\
g_1 &\equiv&  _2F_{1R}(1 - m, 1 - q, 1 - m - q, -1)  e^{-i \pi (m+q)} \; , \\
g_2 &\equiv&  _2F_{1R}(1 + m, 1 + q, 1 + m + q, -1)  \; ,
\eea
\ese
and the $_2F_{1R}$ are regularized hypergeometric functions.

\subsection{Darcx mirror trajectory}

Another analytically simple set of trajectories is found by setting
\be
z(t) = -\f{\xi}{\nu} \sinh^{-1} ( e^{\nu t })\label{darcx-z} \; ,
\ee
where $\nu$ and $\xi$ are constants.  To maintain future asymptotic inertial behavior, it is necessary that
$0< \left| \xi \right| < 1$.  In this case, the mirror begins at rest and
in the limit $t \rightarrow \infty$ its velocity approaches $-\xi$.  We
refer to this set of trajectories as Darcx, short for
\textbf{D}rifting \textbf{Arc}-Hyperbolic Sin of an E\textbf{x}ponential.
A specific example is plotted in Fig.~\ref{fig:trajects}.  Remarkably, the
Bogolubov coefficients $\beta_{\omega \omega'}$ and other relevant quantities
for these trajectories can also be calculated analytically.  Since the
process of deriving them is identical to that discussed previously, we
simply collect the results in Table \ref{darcx-table}.

\begin{table}[t]
\setlength{\extrarowheight}{3pt}
\begin{center}
\caption{Bogolubov coefficients and other useful information for the Darcx
trajectories.}
\label{darcx-table}
\begin{tabularx}{\textwidth}{XX}
\hline
\hline
 &    \\
\noalign{\smallskip}
\hline
{\bf Bogolubov coefficient}     & $\beta_{\w\w'}
= \f{1}{4\pi\sqrt{\w\w'}}\left[-2^{i \w_+ }\f{\xi}{\nu^2}
\f{2\w'\w }{ b_+} \f{\Gamma(-i\w_+)\Gamma(i a_+)}{\Gamma(-i b_+)}\right] $ \\
\quad with $b_{+} \equiv b\w + a\w'$, $a_{+}\equiv a\w + b\w'$,
$\w_{+} \equiv \f{1}{\nu}(\w + \w')$,  &
$|\beta_{\w\w'}|^2  = \f{\xi^2}{4\pi \nu^4} \f{\w'\w}{\w_{+} a_{+} b_{+}}
\f{\text{csch}(\pi\w_{+})\text{csch}(\pi a_{+})}{\text{csch}(\pi b_{+})}$  \\
\quad  $a \equiv \f{1}{2\nu}(1+ \xi)$,
and $b\equiv \f{1}{2\nu}(1-\xi)$.  &   \\
\\
{\bf Proper acceleration}  &  $\alpha(t) = -\frac{\nu \xi e^{\nu t} }
{\left[1- (\xi^2-1) e^{2 \nu t}\right]^{3/2}}$. \\
\\
{\bf Energy flux}  &  $\la T_{uu} \ra = \f{\nu^2}{12\pi}
\f{\xi\chi e^{\nu t_m(u)}[2 (\xi^2-1) e^{2\nu t_m(u)}+1]}
{(\chi + \xi e^{\nu t_m(u)})^2[(\xi^2-1) e^{2\nu t_m(u)} -1]^2}$. \\
\\
{\bf Total energy}  &  $ E_{st} =  \frac{\nu }{96\pi }
\left(\frac{3+\xi ^2 }{2\xi ^2}\ln
\frac{1+\xi }{1-\xi }-\frac{3+\xi  (3+2 \xi )}{ \xi  (1+\xi )}\right)$. \\

\noalign{\smallskip}
\hline
\hline
\end{tabularx}
\end{center}
\end{table}

\subsection{Proex mirror trajectory}

Another interesting trajectory that is asymptotically inertial in the past,
$t \rightarrow -\infty$, can be defined using the Lambert $W$ function,
\be
z(t) = -\f{1}{\rho} W(e^{\rho t}) = -t + \f{1}{\rho}\ln W(e^{\rho t})
= \f{1}{\rho}\ln \left[e^{-\rho t} W(e^{\rho t})\right] \;,
\ee
where the equivalence between the expressions follows from the property
$\ln W(z)=\ln z - W(z)$.  We refer to this trajectory as Proex, which
is short for \textbf{Pro}ductlog \textbf{Ex}ponential.  A plot of its
behavior is overlaid in Fig.~\ref{fig:trajects}.  The late time behavior
of this mirror is similar to the early time behavior of the Carlitz-Willey
trajectory, in that it approaches the speed of light at timelike infinity
while not producing (in this case) a future horizon.  The behavior is best
seen in the Penrose diagram in the right panel of Fig.\ref{fig:trajects}.
Mathematically it can be seen by noting that as
$x\rightarrow \infty$, $W(x) \rightarrow \infty$.  Consequently, the value
of $v$ for the mirror at a given time, $v_m(t)$, has the behavior
$v_m(t) \rightarrow \infty$ as $t \rightarrow \infty$.  The velocity is
\be
\dot{z}(t) = \left[1+W(e^{\rho t})\right]^{-1} - 1 \;,
\ee
which makes obvious the approach to lightspeed as $t\rightarrow \infty$.
The proper acceleration,
\be
\alpha(t) = -\rho \f{ W(e^{\rho t})}{\left[1+2W(e^{\rho t})\right]^{3/2}} \;,
\ee
is initially zero, increases with time until it reaches a maximum magnitude when the
trajectory intersects the null ray $v = 0$, then decreases with time,
vanishing in the limit $t \rightarrow \infty$.

For this trajectory both the ray-tracing function $p(u)$ and its
inverse $f(v)$ can be computed analytically.  The results are
\bes
\bea
\label{proex-p}
p(u) &=& u-\f{1}{\kp}W(2e^{\kp u}) \;,\\
\label{proex-f}
f(v) &=& \f{2}{\kp}e^{\kp v} + v  \;.
\eea
\ese
The Bogolubov coefficients $\beta_{\omega \omega'}$ can in turn be calculated
analytically, as well as the energy flux.  We summarize these and some
other quantities in Table~\ref{proex-table}.

\begin{table}[t]
\setlength{\extrarowheight}{3pt}
\begin{center}
\caption{Bogolubov coefficients and other analytically-derived information
for the Proex trajectory.}
\label{proex-table}
\begin{tabularx}{\textwidth}
{>{\setlength{\hsize}{0.7 \hsize}} X > {\setlength{\hsize}{1.3 \hsize}} X}
\hline
\hline
 &    \\
\noalign{\smallskip}
\hline
{\bf Bogolubov coefficient}
& $ \beta_{\w\w'} =\f{1}{4\pi\sqrt{\w\w'}} \left(\f{2\w'}{\rho}\right)
\left(\f{2\w}{\rho}\right)^{i(\w+\w')/\rho}e^{-\pi (\w+\w')/(2\rho)}
\Gamma\left(-i(\w+\w')/\rho\right) $ \\
 &   $|\beta_{\w\w'}|^2  =  \f{\w'}{2\pi\w\rho(\w+\w')}
\f{1}{e^{2\pi(\w+\w')/\rho}-1}$ \\
 {\bf Proper acceleration}
&  $\alpha(t) = -\rho \f{ W(e^{\rho t})}{[1+2W(e^{\rho t})]^{3/2}}$ \\
{\bf Spectrum of produced particles}
\footnote{Here $\Gamma(0,2\pi \w m/\rho)$ is an upper incomplete gamma
function.} & $\la N_{\w} \ra = -\frac{1}{4\pi^2 \w}\ln\left(1-e^{-2 \pi
\w/\rho}\right) - \frac{1}{2\pi \rho}\sum_{m=1}^{\infty }
\Gamma\left(0,2\pi \w m/\rho\right)$ \\
{\bf Energy flux}
&  $\la T_{uu} \ra =
\f{\rho^2}{48\pi}\f{[2-W(2e^{\rho u})] W(2e^{\rho u})}
{[1+W(2e^{\rho u})]^4}$.
\\
{\bf Total energy}  &  $ E_{st}=\f{\rho}{96\pi} $. \\

\noalign{\smallskip}
\hline
\hline
\end{tabularx}
\end{center}
\end{table}

\subsection{Modified Carlitz-Willey trajectory}

The final class of trajectories that we consider is a modification
of the Carlitz-Willey trajectory.  A term is added that takes the
acceleration to zero at late times.  This causes the trajectory to become
inertial in the future rather than asymptotically null.  The
final velocity is a free parameter.  The trajectories are
\be
z = - \frac{1 - \sigma}{1 + \sigma} \, t
-\frac{1}{\kappa} W \left(\frac{e^{-2 \kappa t/(1+ \sigma)}}{1 +\sigma}\right)
\label{z-mcw}
\;,
\ee
with $0 \le \sigma \le 1$.  A particular example is shown in
Fig.~\ref{fig:trajects} where one can see the divergence from the
Carlitz-Willey mirror at late times.  For $\sigma = 0$ the trajectory
reduces to the Carlitz-Willey one~\eqref{cw-trajectory}.  As before with
the other new trajectories, it proves possible to calculate the Bogolubov
coefficients $\beta_{\omega \omega'}$ analytically, as well as a set of
other relevant quantities.  These are displayed in
Table \ref{mod-CW-table}.

\begin{table}[t]
\setlength{\extrarowheight}{3pt}
\begin{center}
\caption{Bogolubov coefficients and other useful information for the
modified Carlitz-Willey class of trajectories.}
\label{mod-CW-table}
\begin{tabularx}{\textwidth}
{>{\setlength{\hsize}{0.7 \hsize}} X > {\setlength{\hsize}{1.3 \hsize}} X}
\hline
\hline
 &    \\
\noalign{\smallskip}
\hline
{\bf Bogolubov coefficient}     & $ \beta_{\w\w'} =
-\f{1}{2 \kappa \pi} \sqrt{\frac{\w}{\w'}}
 e^{-\pi(\w + \sigma \w')/(2\kp)}
\left(\f{\w'}{\kp}\right)^{-i(\w+\sigma \w')/\kp}
\Gamma\left[i(\w+\sigma \w')/\kp \right]  $ \\
 &   $|\beta_{\w\w'}|^2  = \frac{\w}{2 \pi \kappa \w' (\w + \sigma \w')}
\frac{1}{e^{2 \pi(\w + \sigma \w')/\kappa} - 1}  $ \\
 {\bf Proper acceleration}  &  $\alpha(t) = \frac{\kappa (1+ \sigma)
W\left(e^{-2 \kappa t/(1+\sigma)}/(1+\sigma)\right)}
{2 \left[\sigma + (1+\sigma)
W\left(e^{-2 \kappa t/(1+\sigma)}/(1+\sigma)\right) \right]^{3/2}} $  \\
{\bf Energy flux}  &  $\la T_{uu} \ra =
\frac{\kappa^2}{48 \pi}
\frac{1 - 2 \sigma e^{\kappa u}}{(\sigma e^{\kappa u} + 1)^2 } $. \\

\noalign{\smallskip}
\hline
\hline
\end{tabularx}
\end{center}
\end{table}

\section{Analysis of Energy and Particle Production}
\label{sec:analysis}

For each of the mirror trajectories we consider in this paper, the
renormalized stress-energy for the scalar field is known analytically and
can be easily evaluated.  Except for the Carlitz-Willey trajectory, the
expectation value of the number of particles in a wave
packet, $\la N_{jn} \ra$, cannot be calculated analytically and we
instead evaluate this quantity numerically.  To do so accurately, however,
it proved important to have analytic expressions for the Bogulobov
coefficients $\beta_{\w\w'}$.  The wave packets in turn provide a means of
examining the time and frequency dependence of the created particles.

\subsection{Time dependence}

The correlation (or lack of it) between the number of particles produced
at a given time and the energy flux $\la T_{uu} \ra$
differs markedly from one type of trajectory to the other.  Because the
mirror is in flat space the Bogolubov transformation between the
\textit{in} and \textit{out} vacuum states tells us about the average
number of particles produced in an ensemble of identical systems.  The
energy flux given by the quantity $\la T_{uu}\ra$ gives information about
the average flux of energy produced by the mirror as it accelerates.  The
energy flux is due to a combination of particle production and vacuum
polarization effects.

\begin{figure}[t]
\includegraphics[scale=1.0]{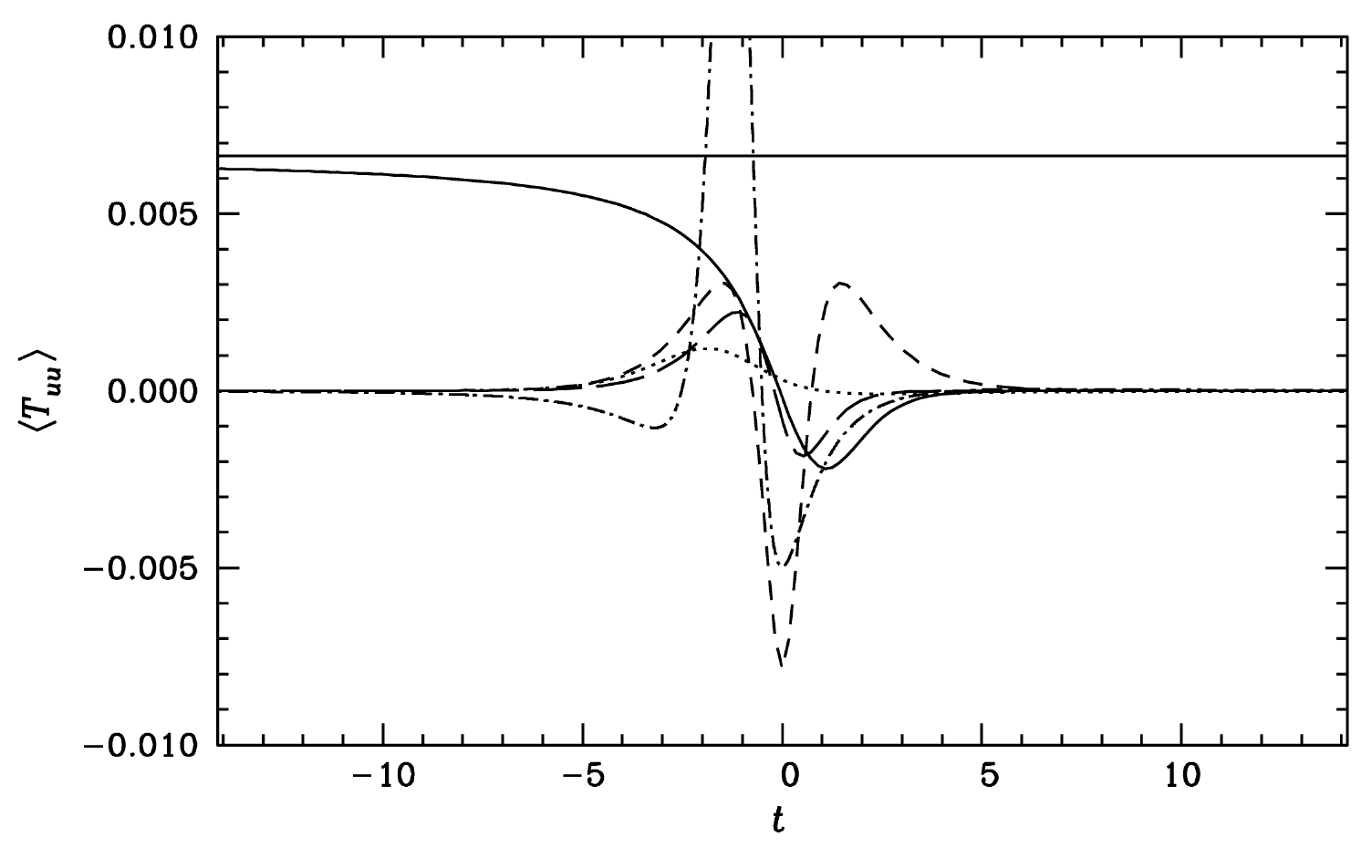}
\caption{\label{fig:time-energyflux} The energy flux
$\langle T_{uu} \rangle$ versus time is plotted for the various mirror
trajectories.  The parameters
$\kappa$, $\mu$, $\nu$, and $\rho$ have all been set equal to $1$.  For
the Darcx trajectory the value $\xi = 1/2$ was chosen and for the modified
Carlitz-Willey trajectory $\sigma = 1/3$ was chosen.  For the Walker-Davies
trajectory $A=2$ and $B=1$ were chosen.  The energy flux in the
Carlitz-Willey case is the constant solid line.  The energy flux associated
with the modified Carlitz-Willey trajectory is the solid curve, which
coincides with the Carlitz-Willey value at early times but then diverges,
briefly resulting in a burst of negative energy before decaying to zero.
The flux associated with the Arctx trajectory is shown as the short dashed
curve and that of the Darcx trajectory is indicated by the long dashed
curve.  The flux from the Proex trajectory is depicted by the dotted curve
and that of the Walker-Davies case by the dot-dashed curve. }
\end{figure}

In Fig.~\ref{fig:time-energyflux} the energy flux is shown for the various
trajectories considered in this paper.  From Eq.~\eqref{stress-alpha} one
can see that the sign of the flux is closely tied with the change in the
proper acceleration of the mirror.  In particular as mentioned in
Sec.~\ref{sec:model} the flux is negative if the change in the proper
acceleration is towards the right and positive otherwise.

\begin{figure}[h]
\includegraphics[scale=1.0]{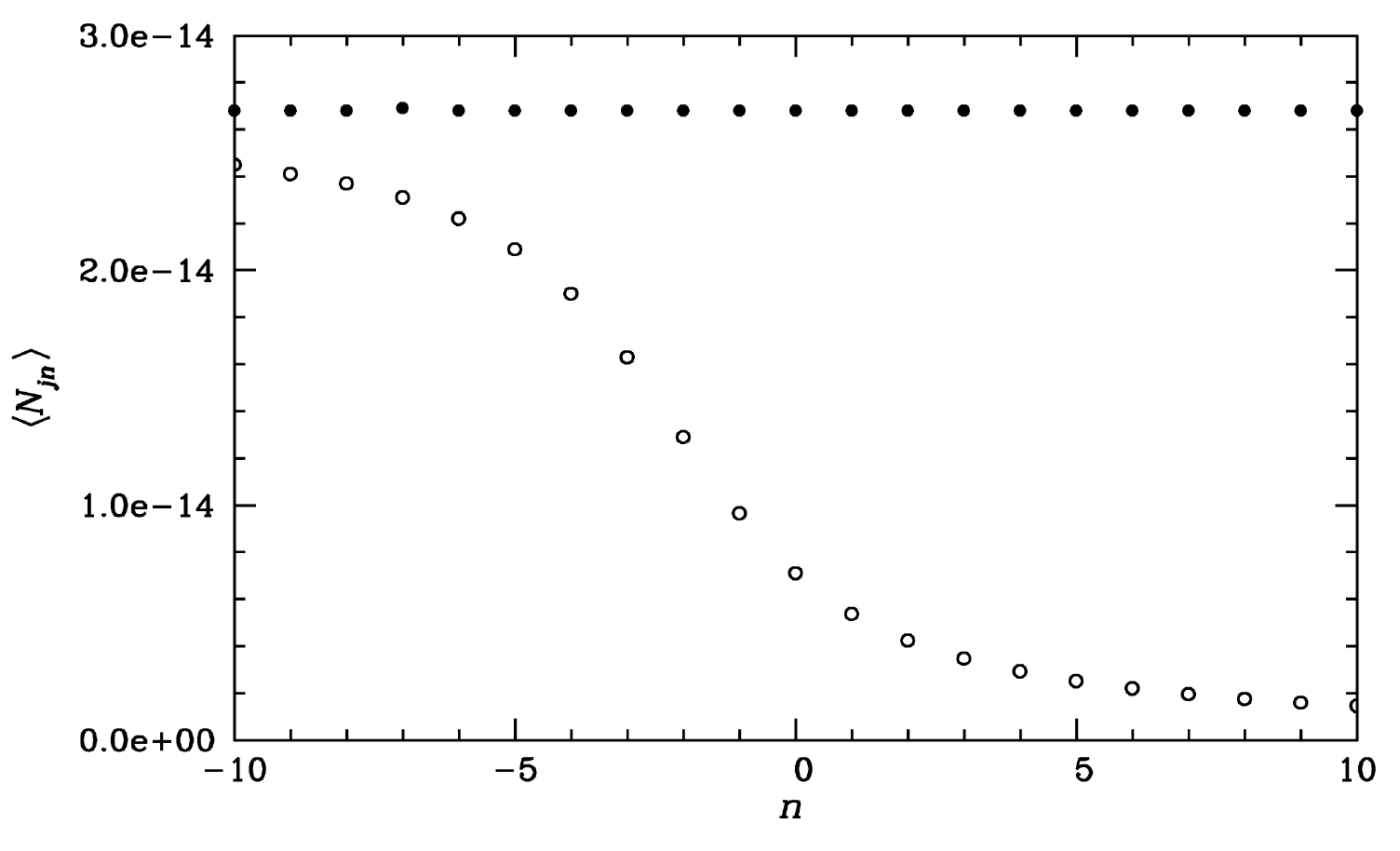}
\caption{\label{fig:time-number-1}
The particle number as measured by wave packets, $\la N_{jn} \ra$, is
plotted as a function of the packet time parameter $n$ for the Carlitz-Willey
(filled circles) and modified Carlitz-Willey (open circles) trajectories.  
In both cases the packet frequency width parameter $\epsilon$ has been 
set to $\sqrt{2} \pi$ and the first non-divergent frequency bin, $j = 1$, 
is shown.  The parameter $\kappa$ has been set to $1$ and for the modified 
Carlitz-Willey trajectory we have taken $\sigma = 1/3$.
}
\end{figure}

\begin{figure}[h]
\includegraphics[scale=1.0]{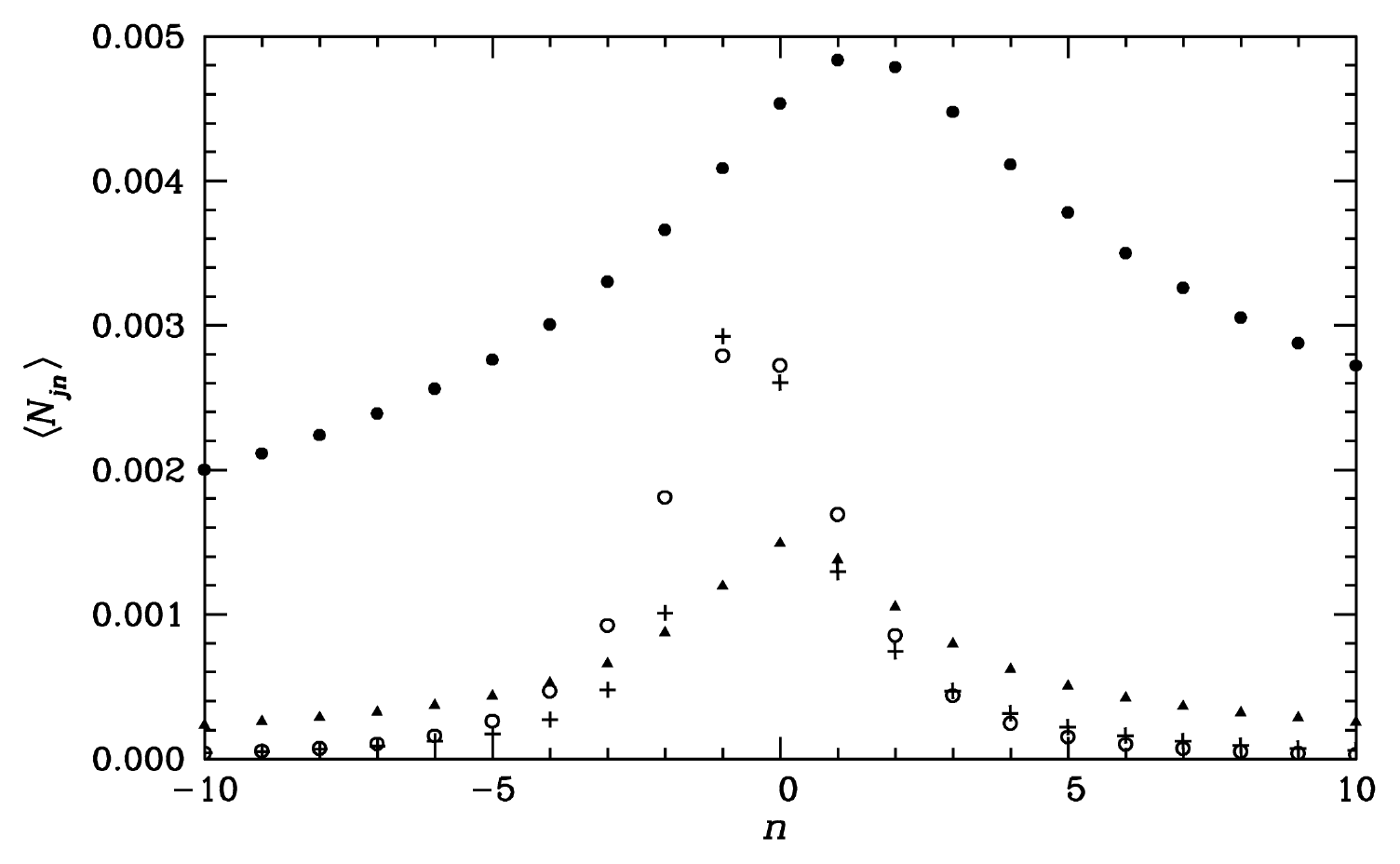}
\caption{\label{fig:time-number-2}
The particle number as measured by wave packets, $\la N_{jn} \ra$, is
plotted as a function of the packet time parameter $n$.  In each case the
packet frequency width parameter $\epsilon$ has been set to $\sqrt{2} \pi$
and the lowest frequency bin $j = 0$ is shown.  The parameters $\mu$, $\nu$,
and $\rho$ have all been set to $1$.  For the Darcx trajectory the value
$\xi = 1/2$ was chosen and for the Walker-Davies case $A=2$ and $B=1$ were
chosen.  The value of $N_{0 n}$ is denoted by the open circles
for the Arctx trajectory, the triangles for the Darcx trajectory,
the filled circles for the Proex trajectory, and the pluses for the
Walker-Davies trajectory.}
\end{figure}

Figs.~\ref{fig:time-number-1} and~\ref{fig:time-number-2} show the
expectation value of the number of particles produced $\la N_{j n} \ra$ as
a function of the time parameter $n$ for the various trajectories we
consider.  In Fig.~\ref{fig:time-number-1}, the frequency parameter $j$ is
set to $1$ because of infrared divergences that occur in $N_{0 n}$ for the
Carlitz-Willey and modified Carlitz-Willey trajectories.  For the
trajectories in Fig.~\ref{fig:time-number-2}, $j$ is set to zero because no
such divergences occur.  The packets of course sample the particle production
discretely.  We draw attention also to the small level of particle excitation 
that occurs for $j=1$ in Fig.~\ref{fig:time-number-1} as compared to that 
which occurs for $j=0$ in Fig.~\ref{fig:time-number-2}.  We return to this 
issue in Sec.~\ref{sec:totalnumber}.

There is a correlation
between the number of particles created during a given time period and the
flux of energy which occurs at that time for the Carlitz-Willey trajectory
in Fig.~\ref{fig:time-number-1} due to the fact that, as discussed in
Sec.~\ref{sec:cw}, the flux is constant in time and the number of particles
$\la N_{j n} \ra$ is independent of the value of the time parameter $n$.
As can be seen in Figs.~\ref{fig:time-energyflux} and~3 a correlation also
occurs for the modified Carlitz-Willey trajectory at early times.  However,
the direct correlation is destroyed by the existence of a negative flux of
energy after the time $t = 0$ when the trajectory has deviated significantly
from the original Carlitz-Willey trajectory.

For the trajectories in Fig.~\ref{fig:time-number-2} there is no direct
correlation between the energy flux and the number of particles created.
In fact for the Arctx trajectory there is something of an anti-correlation
in that at about the time of peak particle production the flux is negative
and has its greatest magnitude.  This shows clearly the limitations in
using the stress-energy tensor to describe the number of particles
created.  Because of vacuum polarization effects, which can include fluxes
of negative energy, it is virtually impossible to separate out the
contribution from the created particles.

\subsection{Frequency spectrum and simultaneous frequency and time resolution}

\begin{figure}
\includegraphics[scale=1.0]{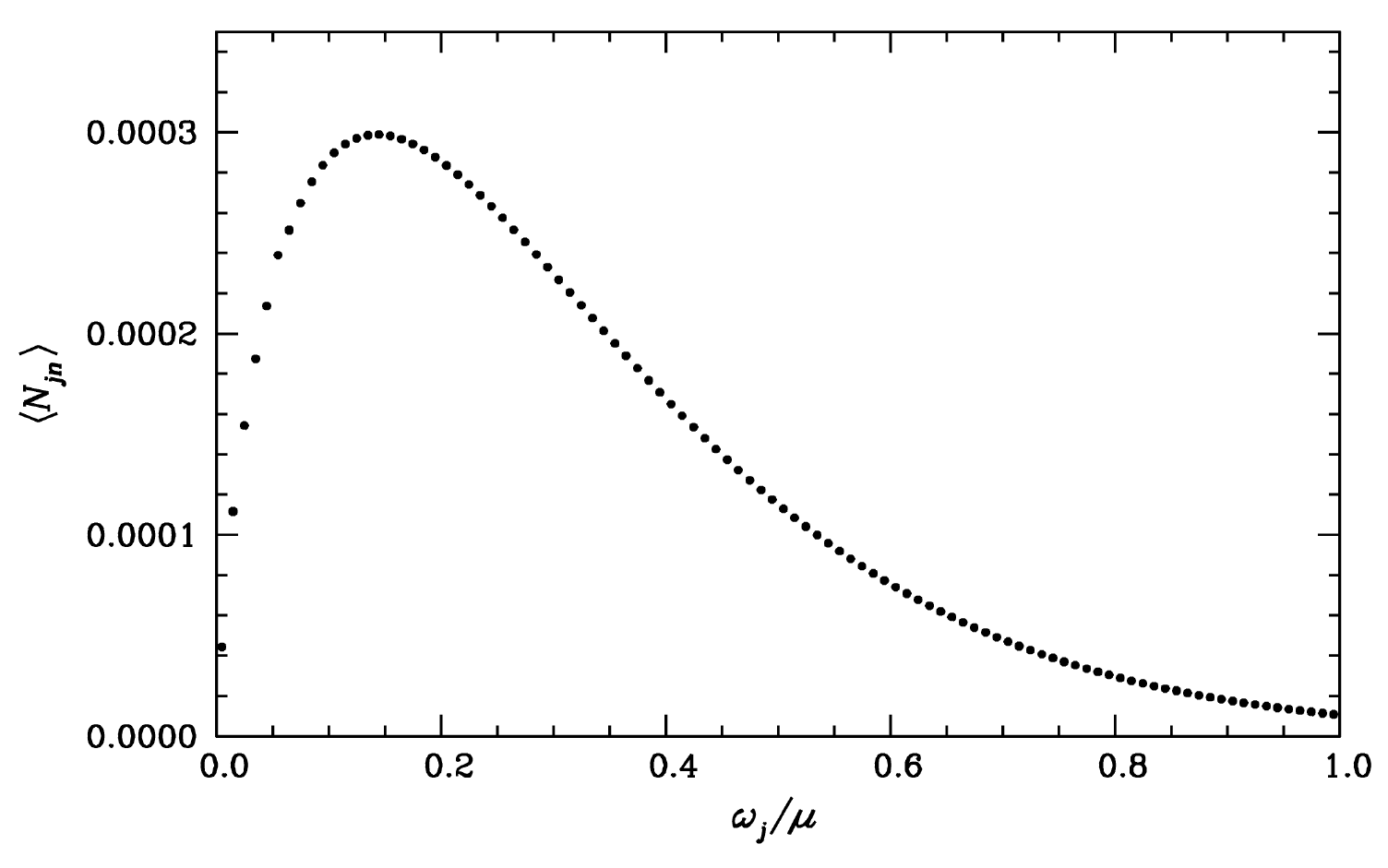}
\caption{\label{fig:cutoffarctxNZERO}
The particle number as measured by wave packets, $\la N_{jn} \ra$, is
plotted as a function of the packet frequency parameter $j$, with $n = 0$
and $\epsilon = 0.01$ for the Arctx trajectory with $\mu = 1$.}
\end{figure}

Wave packets can also be used to measure the frequency spectrum of the
created particles.  Good frequency resolution is obtained by choosing a
small value for $\epsilon$.  As an example consider
Fig.~\ref{fig:cutoffarctxNZERO} for the
$\mu = 1$ Arctx trajectory with a wave packet frequency width parameter
$\epsilon = 0.01$.  The spectrum is a function of packet index $j$, and we
have set $n=0$.  Clearly good frequency resolution is obtained.

In principle the wave packet formalism allows particle production to be
simultaneously resolved, within limits, in both frequency and time.  The
limits, of course, are set by the uncertainty relation.  For any chosen
$\epsilon$, the wave packets have a width in frequency
$\Delta\omega = \epsilon$ and an effective gating interval (width of time
over which the particle detector is on) of $\Delta t \simeq 2 \pi/\epsilon$.
Thus, the time-bandwidth product (for these packets) is $2 \pi$.  It may be
possible to resolve time-dependent spectra for a process with sufficiently
copious particle creation and for which changes in its spectrum occur over
a long enough time interval.  However, for the trajectories studied in this
paper, it has not been possible to resolve the creation process in frequency
and time simultaneously, such that a significant number of particles is
found in many frequency and time bins.  Some of the mirror trajectories have
confined periods during which the acceleration peaks and is strong (i.e.,
Walker-Davies, Arctx, Darcx, and Proex).  For these trajectories,
when we choose a relatively small value of $\epsilon$ to provide good
frequency resolution, we find the vast majority of the particles are created
in the time bin $n=0$.  In contrast, when we choose a relatively large value
of $\epsilon$ to gain good time resolution, we find that almost all of the
particles reside in the lowest frequency bin, $j = 0$.  Furthermore it is
not even possible to find some intermediate value of $\epsilon$ for which
some time and frequency resolution is possible.  Instead, in picking such
an intermediate value of $\epsilon$, we find that the vast majority of
particles lie in the single bin with $n = j = 0$.

A different behavior occurs for the Carlitz-Willey trajectory.  In this case
a small $\epsilon$ will provide adequate frequency resolution and yet not all
of the particle creation occurs in a single time bin, such as $n=0$.  However,
we have not really succeeded in simultaneous time and frequency resolution,
since the Carlitz-Willey mirror gives a spectrum that is completely time
independent.  This then brings us to the modified Carlitz-Willey trajectory.
The energy flux in this case is asymptotically constant in the distant past
but then at some point the energy flux drops toward zero as the acceleration
falls off and the mirror becomes inertial.  The particle creation behaves
similarly.  One might hope that this trajectory would result in a creation
process that could be simultaneously resolved in frequency and time.
However, here too we find that the acceleration falls off sufficiently
rapidly that the transition from creation to effectively no creation
occurs within one time bin (assuming $\epsilon$ has been set to allow good
frequency resolution, i.e. many frequency bins within the transition or
characteristic frequency $\omega_c \sim \kappa$).

To see what seems to be happening, consider the Arctx trajectory.  In that
case the parameter $\mu$ is dimensionally the inverse of time and
ought to represent a characteristic frequency $\omega_c$.  In fact, as
Fig.~\ref{fig:cutoffarctxNZERO} shows, we find the peak of the particle
creation spectrum is $\omega_c \simeq 0.14 \mu$.  However, the period $\tau$
of significant acceleration also depends on $\mu$ and is roughly
$\tau \simeq \mu^{-1}$.  The frequency and time scales are thus related by
a single parameter.  A related factor is that the expectation value of the
total number of particles created $\la N \ra$ is a fraction of unity.  In
other words, in an ensemble of identically-accelerated mirrors, in many
cases there will be no particles produced at all.  Similar time scale and
frequency scale issues occur with the Darcx, Proex, and Walker-Davies
trajectories.

We speculate that a trajectory might be crafted with two parameters: an
acceleration scale $\mu$, with $\tau \simeq \mu^{-1}$, and a duration of
acceleration $T$, which satisfies $T \gg \tau$.  For such a trajectory
enough particles may be created for a long enough period to allow a
significant number of particles to be found in many time and frequency bins,
giving good resolution.  A mirror of this sort would undergo a large change
in Lorentz factor over a time $\sim T$ and it would appear, over that time
interval, like a mirror that is approaching a null horizon.  We have not
been able so far to find a trajectory with these properties for which the
Bogolubov coefficients can be calculated analytically.

The challenge of finding significant wave packet excitation that is
simultaneously spread across a range of both time and frequency bins may
also lie in the inherent nature of the time-bandwidth product of our
orthonormal wave packets.  The time-bandwidth product of these packets is
$2 \pi$, while the fundamental limit of the uncertainty principle is $1/2$.
This may be a contribution to non-uniformity in the creation spectra,
since by their construction, these orthonormal wave packets are unable to
reach the limits of the uncertainty principle.

The fact that it is not possible to obtain significant particle creation
in simultaneous bins of both frequency and time for the trajectories
considered may have interesting
experimental consequences if a system that was in some way like one of
these trajectories could be studied in the laboratory.  However, the
relatively small amount of particle production that occurs might make this
very difficult.

\subsection{Total number of particles produced and their energy}
\label{sec:totalnumber}

In Sec.~\ref{sec:model} expressions were given that allow one to compute
the total number of particles produced and their total energy.  With some
mirror trajectories, we have found one or both of these quantities to be
divergent.

To compute the total number of particles produced one can use
Eq.~\eqref{Ntotal} and to compute their total energy one can use either
Eq.~\eqref{Eqs} or~\eqref{Efromp}.  Comparison of
Eqs.~\eqref{occ},~\eqref{Ntotal}, and~\eqref{Eqs} shows that it is possible
to have a divergent number of particles produced and yet have a finite total
energy, provided that the divergence in the number of particles is due to
an infrared divergence in $N_\omega$ and that it is not too strong.

Substitution of $|\beta_{\w \w'}|^2$ into~\eqref{Ntotal} and~\eqref{Eqs} for 
the Walker-Davies and Arctx trajectories results in both a finite number of 
particles produced and a finite total energy for those particles.  We have 
evaluated $\la N \ra$ numerically for both trajectories.  For the 
Walker-Davies trajectory, $\la N \ra$ depends on the ratio $0 < B/A < 1$.  
As an example, for
$B/A = 1/2$ we find $ \la N \ra  \approx 0.0121$.  For the Arctx trajectory,
$\la N \ra$ is independent of the value of the parameter $\mu$ as can easily
be seen by substituting~\eqref{beta-arctx} into~\eqref{Ntotal} and making
variable changes of the form $\omega = \mu x$.  We find
$\la N \ra \approx 0.0134$.  An interesting point is that both of these
values are small compared to unity.  Thus in an ensemble of identical
systems, in most cases there would be no excitation of any mode and no
particles would be produced.  The total energy produced for the
Walker-Davies trajectory was computed by them and is given in
Eq.~\eqref{Estwalker}.  For the Arctx trajectory it is given in
Eq.~\eqref{E-Tuu-arctx}.  As discussed in Sec.~\ref{sec:model}, the
energies $E_{st}$
in~\eqref{Efromp} and $E_{qs}$ in~\eqref{Eqs} are the same.  This can be
used as a check on the computations of the Bogolubov coefficients
$\beta_{\w\w'}$.  We have computed $E_{qs}$ numerically (in specific
cases for the Walker-Davies trajectories and in general for Arctx) and
found the values to be equal to those for $E_{st}$.

For the other classes of trajectories besides Arctx and Walker-Davies,
the number of particles
$\la N \ra$ diverges.  For the Proex trajectory we have computed
$\la N_\w \ra$ analytically.  Examination of the result, which is shown in
Table \ref{proex-table}, indicates an infrared divergence of the form
\be
\la N_\w \ra \sim - \frac{\ln (2 \pi \omega/\rho)}{\w}  \;.
\ee
This behavior will result in a divergence in the total number of particles
$\la N \ra$.  However when computing $E_{qs}$ in Eq.~\eqref{Eqs} one
multiplies $\la N_\w \ra$ by a factor of $\omega$ before integrating
over $\w$.  The result is a finite value for $E_{qs}$.  We have
computed $E_{qs}$ numerically and found agreement with the expression for
$E_{st}$, which is given in Table \ref{proex-table}.

For the Darcx class of trajectories when $\omega$ is small, one can divide 
the integral in~\eqref{occ} into two parts, one with an integral from $0$ to
$\nu \lambda$ and a second integral from $\nu \lambda$ to $\infty$ with
$1 \gg \lambda \gg \omega/\nu$.  In the first integral one can evaluate
$|\beta_{\w\w'}|^2$ in the limit that both $\omega/\nu$ and $\w'/\nu$ are
small.  The result is a contribution to $\la N_\w \ra$ that is proportional
to $1/\w$.  In the second integral, $\w \ll \w'$ and it is not hard to show
that the integral is finite in the limit $\w \rightarrow 0$.  Thus, as with
the Proex class of trajectories, we find that $\la N \ra$ is infinite but
$E_{qs}$ is finite.  We have numerically computed $E_{qs}$ for specific
values of the parameter $\xi$ and shown that its value is the same as that
for $E_{st}$, which is given in Table \ref{darcx-table}.

For the Carlitz-Willey trajectory, one can see from the form of
$|\beta_{\w \w'}|^2$ that there is both an infrared and ultraviolet
divergence in the integral over $\omega'$ in Eq.~\eqref{occ}.  Thus there
is a divergence in both the particle number and the energy of the produced
particles.  The latter is trivially apparent from the constant flux of energy
which occurs for these trajectories.  For the modified Carlitz-Willey class of
trajectories one can see from the form of $|\beta_{\w \w'}|^2$ in
Table \ref{mod-CW-table} that there is still the infrared divergence but
no ultraviolet divergence when computing $\la N_\omega \ra$.  The infrared
divergence is strong enough to make $\la N_\omega \ra$ divergent for all
values of $\w$.

One might ask what effect the packets have on those trajectories with infrared
divergences.  For the Carlitz-Willey trajectory we were able to analytically
compute the packets and as seen in Eq.~\eqref{Njn-c-w} there is a divergence
for the packets with the lowest frequency range, $j = 0$.  However, for all
other packets $\la N_{j n} \ra$ is finite so the divergences are not nearly
as strong as for $\la N_\w \ra$.

For the modified Carlitz-Willey class of trajectories there is also a
divergence for packets with $j = 0$ but not for those with $j >0$.  To see 
this, one can divide the integral in~\eqref{Njn} into two parts such that 
$I_1 = \int_0^{\kappa \lambda}d \w'$ and 
$I_2  = \int_{\kappa \lambda}^\infty d \w'$, with $0 < \lambda \ll 1$.  For 
the second integral, which contains only nonzero values of $\w'$, it is not
difficult to see that for $\beta_{j n, \w'}$ in Eq.~\eqref{beta-packet} 
there are no infrared divergences resulting from the integral over $\w$.
It can also be seen that $\beta_{j n, \w'}$ is well
enough behaved in the limit $\w' \rightarrow \infty$ that there are no
ultraviolet divergences so long as $\sigma >0$.  For the first integral,
the value of $\w'/\kappa$ is small.  Here it is necessary to break the
discussion into the cases $j = 0$ and $j >0$.  For $j > 0$, we take a
small enough value of $\lambda$ so that $\epsilon > \w'$ for all values of
$\w'$ in the first integral.  Then one can expand the function
$\Gamma[i (\w + \w')/\kappa]$ in the expression for $\beta_{\w \w'}$ in
Table \ref{mod-CW-table} in powers of $\w'/\w$.  Keeping the leading order
term one finds that
\bea
|\beta_{n j, \w'}| &\approx&
\frac{1}{2 \pi \kappa \sqrt{\w' \, \epsilon}} \vline \,
\int_{j \epsilon}^{(j+1)\epsilon} d \w \sqrt{\w} \,
e^{[2 \pi i n/\epsilon - \pi/(2 \kappa)] \w} \;
e^{-i (\w/\kappa) \ln(\w'/\kappa)} \; \Gamma(i \w /\kappa) \, \vline \;.
\label{beta-mcw}
\eea
By repeatedly integrating by parts one can obtain a series in inverse powers 
of $\ln(\w'/\kappa)$.  Substituting into $I_1$ then shows that to leading
order the integrand goes like $1/[\w' (\ln(\w'/\kappa))^2]$, which when
integrated gives no divergence in the limit $\w' \rightarrow 0$.

For $j = 0$ the situation is different.  Here one can divide the integral
in~\eqref{beta-packet} into two parts, $J_1 = \int_0^{\kappa \lambda} d \w$
and $J_2 = \int_{\kappa \lambda}^\epsilon d \w$.  The analysis for the
second integral is exactly the same as for the case $j > 0$.  For the first
integral we choose $\lambda$ to be small enough so that
$ |2 \pi\kappa n/\epsilon - \pi/2| \lambda \ll 1$.  Then
both $\w/\kappa$ and $\w'/\kappa$ are small and 
\be
J_1 \approx \frac{1}{2 \pi \kappa \sqrt{\w' \, \epsilon}} \, \vline
\int_{0}^{\kappa \lambda} d \w \sqrt{\w} \,
e^{-i (\w/\kappa) \ln(\w'/\kappa)} \frac{ \kappa}{\w + \w'} \, \vline \;.
\ee
Making the change of variable $z = -(\w/\kappa) \ln(\w'/\kappa)$, the upper
limit becomes $- \lambda \ln(\w'/\kappa)$, which for fixed $\lambda$ goes
to infinity in the limit $\w' \rightarrow 0$.  The resulting integral can
be computed analytically in terms of Fresnel Integrals.  The result to
leading order in $\w'$ is $J_1 \sim 1/\sqrt{- \w' \ln(\w'/\kappa)}$.  This
gives a contribution to $I_1$ that when integrated results in a divergence
at the lower limit $\w' = 0$ and thus a divergence
in $\la N_{j n} \ra$ for $j = 0$.

For the Darcx and Proex trajectories it turns out there is no infrared
divergence in $\la N_{j n} \ra$ even for $j = 0$ and even though
$\la N_\w \ra$ diverges in the limit $\w \rightarrow 0$.  This can be shown
by bounding $\la N_{jn} \ra$ by
substituting Eq.~\eqref{beta-packet} into Eq.~\eqref{Njn} and then computing
the absolute values of each factor in the integrand of~\eqref{beta-packet}.
The result is
\be
\la N_{j n} \ra \le \int_0^\infty d \w'
\left(\int_{j \epsilon}^{(j+1)\epsilon} d \w |\beta_{\w\w'}| \right)^2 \;.
\ee
It is not hard to show that for small $\w'$ the integrand is finite in the
limit $\w' \rightarrow 0$ if $j > 0$ for both trajectories.  If $j = 0$ then
it is useful to again divide the integral over $\w$ into two parts as was
done above, except with $\kappa \rightarrow \nu$ and $\rho$ respectively
for the Darcx and Proex trajectories.  Then as before the
analysis for $J_2$ is the same as for $j > 0$.  For $J_1$ both $\w$ and
$\w'$ are small making it possible to expand the terms in $|\beta_{\w\w'}|$.
When this is done and the integral over $\w$ is computed for the leading
order terms, we find that the result is finite in the limit
$\w' \rightarrow 0$ for the Darcx trajectories.  For the Proex trajectory,
we find that to leading order $J_1 \sim \ln \w' $ so that the integrand
for the integral over $\w'$ goes like $(\ln \w')^2$ and a finite
contribution is made to $\la N_{0n} \ra$.

As discussed in Sec.~\ref{sec:model}, one can obtain an estimate of the
energy of the created particles using the packet formalism by
multiplying $\la N_{jn} \ra$ by the frequency in the middle of the range
for each packet and summing over $j$ and $n$ as in Eq.~\eqref{Eep}.
The resulting energy, $E_{ep}$, has been computed for the Arctx
trajectory for two different values of $\epsilon$.  For the case shown
in Fig.~\ref{fig:cutoffarctxNZERO} with $\epsilon = 0.01$ the results
agreed with $E_{st}$ in Eq.~\eqref{E-Tuu-arctx} to within about
$0.01\%$.  In a separate calculation, with $\epsilon = 10$, the agreement
was at the $1\%$ level, which is remarkably good given the poor
frequency resolution, which might be expected to drastically skew the
energy summation.

For the Darcx class of trajectories with $\xi = 0.99$, $\sigma = 1$, and
$\epsilon = 0.001$, we find agreement with the values of $E_{st}$ to within
four digits.  The result was obtained by summing packets with $n = 0$ and
values of $j$ ranging from $j = 0$ to $j=1175$.

For the Proex trajectory
with $\rho = 1$, a numerical computation of $E_{ep}$ for
$\epsilon = 5 \times 10^{-5}$ gave results in agreement with $E_{st}$ to
approximately $0.08\%$.  Energy packets with $n=0$ and with a sum from
$j = 0$ to $j = 12000$ were used in that calculation.

\section{CONCLUSION}
\label{sec:conclusions}

We have investigated the particle production and the energy flux that results
from a massless, minimally coupled scalar field in a two-dimensional flat
space that contains an accelerating mirror.  Dirichlet boundary conditions
are assumed at the mirror and the field is assumed to be in the \textit{in}
vacuum state.  Six different types of trajectories have been considered,
including
the one studied previously by Carlitz and Willey~\cite{Carlitz:1986nh} and
the one studied by Walker and Davies~\cite{Walker:1982}.  The other four
are new and have been introduced for this study.  These trajectories are
all asymptotically inertial in the limit $t \rightarrow -\infty$ and all but
the Carlitz-Willey trajectory are also asymptotically inertial in the
limit $t \rightarrow \infty$.

For each trajectory it has been possible to obtain analytically the Bogolubov
coefficients $\beta_{\w \w'}$ as well as the proper acceleration and the
energy flux $\la T_{uu} \ra$.  As pointed out by Walker~\cite{Walker:1984vj},
it is very useful to have models in which it is possible to do analytic
calculations.  The four new types of trajectories that we have provided
fit this description.

Our main focus has been on the use of wave packets, which allow the particle
production to be time resolved, and in principle might allow significant
simultaneous
frequency resolution as well.  The packets we use form a complete orthonormal
set, so that no information is lost.  By computing the Bogolubov coefficients
for the packets and integrating over the frequencies of the \textit{in}
modes, it is possible to obtain the average number of particles reaching
$\mathscr{I}^+$ found in a given frequency range and an approximate time
range (more specifically a range in the null coordinate $u$).  Thus this
method of analysis can be thought of as similar to what a series of particle
detectors along a large $v$ surface would detect if each was turned on for
some relatively short period of time.

In principle one might expect a correlation between the time dependence of the
particle production and the energy flux $\la T_{uu} \ra$.  However both vacuum
polarization and particle production effects are combined in the
stress-energy tensor and it is difficult if not impossible to separate
them.  Use of the wave packet formalism allows for an unambiguous
description of the time dependence of the particle production process.

The Carlitz-Willey trajectory was designed to result in a constant flux of
energy and is of a different nature than the other trajectories in being
asymptotically null.  We found an explicit mathematical expression for this
trajectory in terms of the Lambert $W$ function.  Because of the constant
flux of energy the total energy produced is divergent.  Not surprisingly
it is also found that the number of particles
produced per frequency interval $\la N_\w \ra$ is also divergent.

This trajectory was the only one for which we were able to compute the
number $\la N_{jn}\ra$ of particles associated with a wave packet
analytically.  Just as the energy flux is constant, we found that
$\la N_{jn} \ra$ is independent of the value of $n$ and so is time
independent.  A divergence occurs for the case $j = 0$ but not for larger
values of $j$.  This infinity can be dealt with by simply ignoring
the lowest frequency ($j=0$) bin.  In a realistic detector there will
always be an infrared cutoff, since it is impossible to detect particles
of arbitrarily long wavelengths.  An exploration of the effects on the
frequency range for the wave packets with $j >0$ was carried out and
it was found that a Planck type spectrum is approached in the limit that
the frequency width of the packets vanishes.

The modified Carlitz-Willey trajectory has the same approximate behavior as
the Carlitz-Willey trajectory at early times but then stops accelerating and
approaches a constant velocity at late times.  As with the Carlitz-Willey
trajectory the number of particles produced per frequency interval
$\la N_\w \ra$ is divergent for all $\w$.  The energy flux $\la T_{uu} \ra$
is approximately constant at early times and this results in an infinite
amount of total energy $E_{st}$.  Using wave packets we again find that
$\la N_{jn} \ra$ is divergent for $j = 0$ but finite for all other values
of $j$.  For $j > 0$ we find that, as a function of the time parameter
$n$, $\la N_{jn} \ra$ is approximately constant at early times and
decreases to zero at late times as would be expected for a trajectory
that is asymptotically inertial.

The Arctx and Walker-Davies trajectories are the only ones for which the
mirror begins and ends at rest.  We find a finite number of particles
$\la N \ra$ is produced for both along with a finite amount of energy
$E_{st}$.  Using wave packets we have shown that the number of particles
produced $\la N_{j n} \ra$  increases to a maximum and then decreases over
the range of time that the mirror's acceleration is first increasing and
then decreasing at a significant rate.

For the Darcx and Proex trajectories the mirror begins at rest and is
asymptotically inertial in the future.  In the Darcx case it approaches a
constant speed that is less than that of light and in the Proex case it
approaches the speed of light but in such a way that it is not asymptotic
to a null trajectory.  For these trajectories there is
an infrared divergence in the number of particles produced but not in the
energy of the produced particles.  However there is no divergence in
$\la N_{j n} \ra$ for $j = 0$.  We find that the number of particles first
increases to a maximum and then decreases during the period when the
acceleration is first increasing and then decreasing at a significant rate.

It is interesting to compare the results for the time-dependent particle
production with the average energy flux $\la T_{uu}\ra$.  Not surprisingly,
for the Carlitz-Willey trajectory both are constant in time, so in that
sense there is a correlation.  For the modified Carlitz-Willey class of
trajectories there is a similar correlation at early times.  Once the
number of particles produced begins decreasing, the correlation diminishes
and even disappears due to a flux of negative energy that occurs at
intermediate times.  For the other trajectories, probably again because of
the fluxes of negative energy during certain time periods, there is little
or no correlation between the number of particles produced and the energy
flux.

Because the wave packets tile both the time and frequency domains, it might
be expected that one could obtain time-dependent spectra for the particle
production.  For each type of trajectory (except Carlitz-Willey) the time
and frequency ranges of the bins were varied.  It was found (at least for
these trajectories) that simultaneous time and frequency resolution with
significant particle build-up in each domain is absent.  The mirrors have
a single dimensional parameter that determines both the characteristic
frequency and duration of creation, and for this reason the uncertainty
principle prevents one from measuring the spectral dynamics.  It may be
possible to find mirror trajectories with two characteristic scales, one
that sets the acceleration and characteristic frequency and one that sets
a duration of creation.  This would allow significant particle creation in
the spectral dynamics to be measured by wave packets.  We have not found
such a trajectory.  Conversely, it may be a generic feature of the quantum
nature of accelerating mirrors.  Our results are in two dimensions, not
four.  However, it seems unlikely that this effect of the uncertainty
principle is tied to the number of dimensions.  If our results are pointing
to a generic effect, then it could have important observational consequences
for any experiments that might attempt to detect the radiation produced when
a mirror accelerates.

\begin{acknowledgments}

M.R.R.G. appreciates helpful discussions with Xiong Chi, Paul Davies, Adam
Kelleher, Laura Mersini-Houghton, and Ryan Rohm.  P.R.A acknowledges helpful
discussions with Jason Bates and Sarah Fisher.  M.R.R.G. acknowledges
support from the US Department of Education GAANN Fellowship Program Grant
Number P200A090135.  C.R.E. acknowledges support from the Bahnson Fund at
the University of North Carolina--Chapel Hill.  This work was supported in
part by the National Science Foundation under grant numbers PHY-0556292 and
PHY-0856050 to Wake Forest University.

\end{acknowledgments}

%\bibliographystyle{unsrt}   % this means that the order of references
			    % is determined by the order in which the
			    % \cite and \nocite commands appear
\bibliography{mirror}  % list here all the bibliographies that
			     % you need.

\end{document}